\documentclass{aastex631}
\usepackage[T1]{fontenc}
\usepackage{epsf,color,amsmath}

\usepackage{cancel}

\newcommand{\sfrac}[2]{\mathchoice%
  {\kern0em\raise.5ex\hbox{\the\scriptfont0 #1}\kern-.15em/
    \kern-.15em\lower.25ex\hbox{\the\scriptfont0 #2}}
  {\kern0em\raise.5ex\hbox{\the\scriptfont0 #1}\kern-.15em/
    \kern-.15em\lower.25ex\hbox{\the\scriptfont0 #2}}
  {\kern0em\raise.5ex\hbox{\the\scriptscriptfont0 #1}\kern-.2em/
    \kern-.15em\lower.25ex\hbox{\the\scriptscriptfont0 #2}} {#1\!/#2}}

\newcommand{\castro}{{\sf Castro}}
\newcommand{\maestroex}{{\sf MAESTROeX}}
\newcommand{\pynucastro}{{\sf pynucastro}}

\newcommand{\amrex}{{\sf AMReX}}

\newcommand{\isot}[2]{$^{#2}\mathrm{#1}$}
\newcommand{\isotm}[2]{{}^{#2}\mathrm{#1}}

\newcommand{\omegadot}{\dot{\omega}}
\newcommand{\Sdot}{\dot{S}}
\newcommand{\ddx}[1]{{\frac{{\partial#1}}{\partial x}}}

\newcommand{\ddt}[1]{{\frac{{\partial#1}}{\partial t}}}
\newcommand{\odt}[1]{{\frac{{d#1}}{dt}}}

\newcommand{\dedXd}{\left . \frac{\partial{}e}{\partial{}X_k} \right |_{\rho, T, X_{j,j\ne k}}}

\newcommand{\Ic}{{\boldsymbol{\mathcal{I}}}}

\newcommand{\kmax}{{k_\mathrm{max}}}

\usepackage{bm}

\newcommand{\Uc}{{\,\bm{\mathcal{U}}}}
\newcommand{\Fb}{\mathbf{F}}
\newcommand{\Sc}{\mathbf{S}}

\newcommand{\xv}{{(x)}}

\newcommand{\Ab}{{\bf A}}
\newcommand{\Sq}{{\bf S}_\qb}

\newcommand{\qb}{{\bf q}}

\newcommand{\Gb}{{\bf G}}
\newcommand{\Rb}{{\bf R}}

\newcommand{\Adv}[1]{{\left [\boldsymbol{\mathcal{A}} \left(#1\right)\right]}}
\newcommand{\Advt}[1]{{\left [\boldsymbol{\mathcal{\tilde{A}}} \left(#1\right)\right]}}

\newcommand{\Advs}[1]{\boldsymbol{\mathcal{A}} \left(#1\right)}

\newcommand{\atolspec}{{\epsilon_\mathrm{abs,spec}}}
\newcommand{\rtolspec}{{\epsilon_\mathrm{rel,spec}}}
\newcommand{\atolener}{{\epsilon_\mathrm{abs,ener}}}
\newcommand{\rtolener}{{\epsilon_\mathrm{rel,ener}}}

\begin{document}
%======================================================================
% Title
%======================================================================
%\title{A Simplified Spectral Deferred Correction Method for Coupling Hydrodynamics with Reaction Networks and Nuclear Statistical Equilibrium}
\title{An Improved Method for Coupling Hydrodynamics with Astrophysical Reaction Networks}

%\shorttitle{A Simplified SDC Method}
\shorttitle{Hydro/Reaction Coupling with NSE}

\shortauthors{}

\author[0000-0001-8401-030X]{M.~Zingale}
\affiliation{Dept.\ of Physics and Astronomy, Stony Brook University,
  Stony Brook, NY 11794-3800}

\author[0000-0003-0439-4556]{M.~P.~Katz}
\affiliation{Dept.\ of Physics and Astronomy, Stony Brook University,
  Stony Brook, NY 11794-3800}

\author[0000-0003-1791-0265]{A.~Nonaka}
\affiliation{Center for Computational Sciences and Engineering, Lawrence Berkeley National Laboratory, Berkeley, CA  94720}

\author[0000-0002-0297-0313]{M.~Rasmussen}
\affiliation{Utah State University, Logan UT 84322}

\correspondingauthor{Michael Zingale}
\email{michael.zingale@stonybrook.edu}

%======================================================================
% Abstract and Keywords
%======================================================================
\begin{abstract}
Reacting astrophysical flows can be challenging to model because of
the difficulty in accurately coupling hydrodynamics and reactions.
This can be particularly acute during explosive burning or at high
temperatures where nuclear statistical equilibrium is established.  We
develop a new approach based on the ideas of spectral deferred
corrections (SDC) coupling of explicit hydrodynamics and stiff
reaction sources as an alternative to operator splitting that is simpler
than the more
comprehensive SDC approach we demonstrated previously.  We apply the
new method to a double detonation problem with a moderately-sized
astrophysical nuclear reaction network and explore the timestep size
and reaction network tolerances to show that the simplified-SDC
approach provides improved coupling with decreased computational
expense compared to traditional Strang operator
splitting.  This is all done in the framework of the
\castro\ hydrodynamics code, and all algorithm implementations are
freely available.
\end{abstract}


%======================================================================
% Introduction
%======================================================================
\section{Introduction}\label{Sec:Introduction}

Modeling astrophysical reacting flows can be challenging because of
the disparity between the nuclear and hydrodynamics timescales.
Reaction networks tend to be stiff, requiring implicit integration
techniques to stably integrate the system~\citep{BYRNE19871}.  In contrast, compressible
hydrodynamics flows are limited by the (often much longer)
sound-crossing time over a computational cell and can be solved using
explicit time integration. Traditional methods of coupling
hydrodynamics and reactions used in astrophysics use operator
splitting---each physical process acts on the output of the previous
process in alternating fashion.  This makes it easy to use different
time-integration methods for the different physics, and to build a
simulation code in a modular way.  However, competition between the
different physical processes can cause the coupling to break down, since the reactions do not directly incorporate the effects of the hydrodynamics and vice versa.
These splitting errors can lead to loss of accuracy and further time
step limitations.

A particularly difficult phase of evolution to model is the nuclear
statistical equilibrium that sets in for temperatures in excess of
$\mbox{few} \times 10^9~\mathrm{K}$.  Physically, the forward and
reverse rates of reaction should balance leading to an equilibrium.
With operator splitting, an NSE region will have a large positive flow
through the network in a zone in one step followed by a large negative
flow over the next timestep, as the code struggles to produce an
equilibrium. 
These large changes in abundances (and large alternately positive and negative energy generation rates) can be a challenge for
the integration method---it may take more steps than allowed by the ODE integrator, require a timestep below floating point accuracy, or fail to meet the tolerances.  The easiest way to improve the coupling is to cut the
timestep (see, e.g., \citealt{couch:2015,rivas:2022}), but this can make simulations
prohibitively expensive.  Sometimes the burning is simply halted on a
zone-by-zone basis when NSE conditions are reached (e.g., as in
\citealt{hedet}).  The
present work focuses on networks alone, but in a follow-on paper we
will explore hybrid burning models consisting of networks and tables for
nuclear statistical equilibrium.

The \castro\ hydrodynamics code \citep{castro,castro_joss} is used for
all of our numerical experiments.  \castro\ is a compressible
(magneto-, radiation-) hydrodynamics code built on the AMReX adaptive
mesh refinement (AMR) framework \citep{amrex_joss}.  \castro\ has been
designed to be performance portable and runs on massively parallel CPU,
multicore, and GPU architectures \citep{castro_gpu}.  For hydrodynamics, the
corner transport upwind (CTU; \citealt{ppmunsplit}) method with the
piecewise parabolic method (PPM; \citealt{ppm,millercolella:2002}) is
used.  \castro\ includes self-gravity, rotation, arbitrary equations
of state and reaction networks, and has been used for modeling X-ray
bursts and different models of thermonuclear, core-collapse, and
pair-instability supernovae.  Recently in \citet{castro_sdc}, we
developed second- and fourth-order accurate in space and time
method-of-lines approaches for coupling hydrodynamics and nuclear reaction networks based on
spectral deferred corrections (SDC), and demonstrated these
methods using a variety of test problems.

The time-integration approach presented here is considerably simpler
than the SDC method of \citet{castro_sdc}, but allows us to reuse the
piecewise-linear or piecewise-parabolic CTU hydrodynamics
construction \citep{saltzman1994,millercolella:2002}
used in the the original \castro\ paper
%\MarginPar{any cites to CCSE papers? - you would want Colella90 (2D CTU), Saltzman94 (3D CTU), Miller/Colella02 (2D/3D PPM)}
and a largely similar ODE integration scheme, making this
method easier to add to existing simulation codes.  Furthermore, it
also extends to adaptive mesh refinement with subcycling in a
straightforward manner, avoiding the complications described in
\cite{mccorquodalecolella} needed to fill ghost cells when using
method-of-lines integration.  However, it is restricted to
second-order accuracy in time overall.  We term this algorithm the
``simplified-SDC method''.  In this paper we describe the overall
method and demonstrate it on a test problem and astrophysical
simulation using \castro.  All of the code to reproduce the results in
this paper are in the \castro\ GitHub
repository\footnote{\url{https://github.com/amrex-astro/Castro/}}.

\section{Numerical Methodology}

We solve the Euler equations for compressible, reacting flow.  For ease
of exposition we describe the one-dimensional case;
multidimensional extensions are a straightforward modification to
include in the CTU hydrodynamics scheme.  Our conserved variables are
\begin{equation}
  \Uc = \left ( \begin{array}{c}
           \rho \\
           \rho X_k \\
           \rho u \\
           \rho E \\
           \rho e \end{array}\right )
\end{equation}
where $\rho$ is the mass density, $u$ is velocity, $E$ is specific
total energy, $p$ is the pressure, and we carry nuclear species mass
fractions, $X_k$.  The specific total
energy relates to the specific internal energy, $e$, as $E = e + u^2/2$,
and we also separately evolve $e$, as part of a dual
energy formulation (see \citealt{bryan:1995,wdmergerI}).
 The mass fractions are constrained to sum to 1, $\sum_k X_k = 1$.
Defining the hydrodynamical fluxes:
\begin{equation}
  \Fb(\Uc) = \left ( \begin{array}{c}
         \rho u \\
         \rho X_k u \\
         \rho u^2 p \\
         (\rho E + p) u \\
         \rho e u \end{array}\right )
\end{equation}
We can write the system in conservative form for all state variables aside from $(\rho e)$ as:
\begin{equation}
  \ddt{\Uc} + \ddx{\Fb(\Uc)} = \Sc(\Uc)
\end{equation}
where for the special case of $(\rho e)$, we have an additional ``$p\, dV$'' term:
\begin{equation}
\ddt{(\rho e)} + \ddx{F_{\rho e}} + p \ddx{u} = S_{\rho e}
\end{equation}
We split the source term into gravitational and reactive parts, $\Sc(\Uc) = \Gb(\Uc) + \Rb(\Uc)$, with
\begin{equation}
  \label{eq:cons_sources}
  \Gb(\Uc) = \left ( \begin{array}{c}
    0 \\
    0 \\
    \rho g \\
    \rho u g \\
    0 \end{array} \right ),
  \qquad
  \Rb(\Uc) = \left ( \begin{array}{c}
     0 \\
     \rho \dot\omega_{X_k} \\
     0 \\
     \rho \dot{S} \\
     \rho \dot{S}
  \end{array} \right ).
\end{equation}
From reactions,
$\dot\omega_{X_k}$ is the creation rate for species $k$
 and $\dot{S}$ is the energy generation rate per unit
mass.  We note that the internal energy $p\partial u/\partial x$ (``$p\, dV$'')
work is not treated as
a source term but is instead constructed with the hydrodynamical
fluxes that are computed in the CTU method.
Aside from the ``$p\, dV$'' term in the internal
energy equation, this system is in conservative form with source
terms.

We define the advective terms with gravitational sources, $\Advs{\Uc}$, as
\begin{equation}
\Advs{\Uc} = -\ddx{\Fb(\Uc)} + \Gb(\Uc)
\end{equation}
for the general case, with the $(\rho e)$ component again having the extra
``$p\, dV$'' term:
\begin{equation}
\mathcal{A}_{\rho e} = -\ddx{F_{\rho e}} -p \ddx{u} + G_{\rho e}
\end{equation}
To close the system, we need an equation of state of the form:
\begin{equation}
p = p(\rho, {\bf X}, e).
\end{equation}
Sometimes it is preferable to work with the primitive variables:
\begin{equation}
\qb = \left ( \begin{array}{c}
  \rho \\
  X_k \\
  u \\
  p \\
  (\rho e) \\
\end{array} \right )
\end{equation}
Here, the system appears
as:
\begin{equation}
\qb_t + \Ab^\xv(\qb) \qb_x  = \Sc(\qb)
\end{equation}
with the matrix $\Ab^\xv$ giving the coefficients of the spatial derivatives
of the primitive variables:
\begin{equation}
\Ab^\xv(\qb) = \left ( \begin{array}{ccccc}
    u & 0 &  \rho & 0 & 0 \\
    0 & u &  0    & 0 & 0 \\
    0 & 0 &  u    & 1/\rho & 0 \\
    0 & 0 &  \Gamma_1 p & u & 0 \\
    0 & 0 &  \rho h & 0 & u
  \end{array} \right )
\end{equation}
where $h$ is the specific enthalpy and $\Gamma_1$ is an adiabatic index,
$\Gamma_1 = d\log p/d\log\rho|_s$ at constant entropy.
The CTU+PPM algorithm uses the characteristic wave structure of $\Ab$
to collect the information that makes it to an interface over a timestep
in order to compute the fluxes through the interface.
Note, the primitive state has two thermodynamic quantities, $p$
and $(\rho e)$, to more efficiently handle the general equation of
state in the Riemann solver, as described in \citet{castro}, but
alternate formulations are possible \citep{colellaglaz:1985}.
The source term vector, $\Sc(\qb)$, can again be decomposed into gravitational
sources (now in terms of the primitive variables) and reaction terms,
\begin{equation}
  \Sc(\qb) = \Gb(\qb) + \Rb(\qb),
\end{equation}
with
\begin{equation}
\label{eq:prim_sources}
\Gb(\qb) = \left ( \begin{array}{c}
     0 \\
     0 \\
     g \\
     0 \\
     0 \\
   \end{array} \right ),
\qquad
\Rb(\qb) = \left ( \begin{array}{c}
     0 \\
     \dot\omega_{X_k} \\
     0 \\
     \Gamma_1 p \sigma \Sdot \\
     \rho \Sdot
   \end{array} \right ),
\end{equation}
where
\begin{equation}
\sigma \equiv \frac{\partial p/\partial T |_\rho}{\rho c_p \partial p/\partial \rho |_T}
\end{equation}
and $c_p$ is the specific heat at constant pressure, $c_p = \partial
h/\partial T |_p$.  A derivation of this source for the pressure
equation can be found in \cite{ABNZ:III}.  We note that this source is
algebraically identical to that shown in Equation (25) of \cite{castro}.

The CTU+PPM method for hydrodynamics is second-order accurate in space
and time.  We want to couple the reactive sources to the hydrodynamics
to be second-order in time as well.  As discussed above, nuclear
reaction sources are stiff, and need to be integrated using implicit
methods for stabilty.  Operator splitting (e.g., Strang) is
traditionally employed here, and is used as a benchmark for comparison
in this paper.  We discuss this traditional approach next before moving
on to our new time-coupling method.

\subsection{Strang Splitting}

\begin{figure}[t]
  \plotone{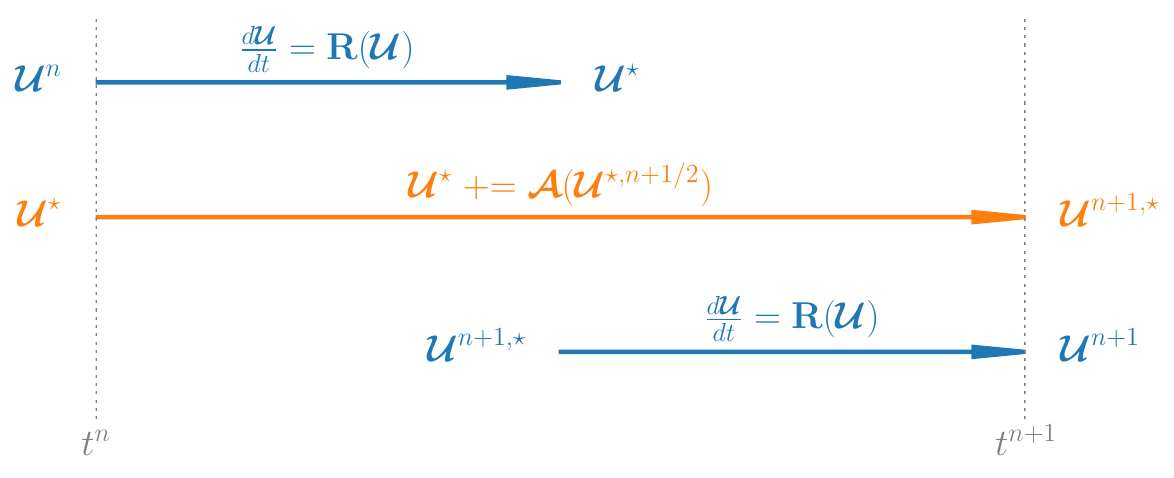}
  \caption{\label{fig:strang_flowchart} Flowchart of the operator split (Strang) time integration.  We first react for $\Delta t/2$, then
    advect for $\Delta t$, and finally react for $\Delta t/2$.  Each
    process uses the state left behind from the previous process.}
  \end{figure}

In the Strang splitting~\citep{strang:1968} flavor of operator
splitting, we first integrate the system with reactions terms only (no
advection) over $\Delta t/2$, then integrate the advection terms only
(no reactions) over $\Delta t$, and finally integrate the reaction
terms only over $\Delta t/2$.  The staggering of the reactive update
means the hydro implicitly sees a time-centered reactive source, making
the update second-order accurate in time.  This flow is illustrated in
Figure~\ref{fig:strang_flowchart}.

In the absence of advective terms, our reaction system appears as just
$d\Uc/dt = \Rb(\Uc)$, or:
\begin{align}
\odt{\rho} & = 0 \\
\odt{(\rho X_k)} &= \rho \dot\omega_{X_k} \\
\odt{(\rho u)} &= 0 \\
\odt{(\rho E)} &= \rho \Sdot
\end{align}
Notice that in the Strang formulation density is held constant when reacting.
We can write the energy equation as:
\begin{equation}
\odt{(\rho E)} = \odt{(\rho e)} + \odt{K} = \rho \Sdot
\end{equation}
where $K$ is the kinetic energy, $K = \rho |u|^2/2$.  Since the density and velocity
are unchanged by reactions (when Strang splitting), our energy equation becomes:
\begin{equation}
\label{eq:strang:e}
\odt{(\rho e)} = \rho \odt{e} = \rho \Sdot
\end{equation}
The reaction rates are typically expressed as $\omegadot_k(\rho, T, {\bf X})$
when we evolve this system, which requires us to get temperature from
the equation of state each
time we need to evaluate the reactive terms.

We also typically integrate mass fractions, instead of partial
densities:
\begin{equation}
\label{eq:strang:X}
\odt{X_k} = \omegadot_k
\end{equation}
We integrate Equations (\ref{eq:strang:e}) and (\ref{eq:strang:X})
using an implicit ODE solver described below.

We explored this and other approaches (including not evolving an
energy / temperature equation during the reaction step) in
\citet{strang_rnaas} and showed that the above formulation achieved second
order convergence and that in particular, integrating an energy / temperature
equation is needed to get second order accuracy (see also
\citealt{muller:1986} for a discussion on stability).  We note however
that some astrophysical simulation codes only evolve the species
equations.  For very strong reactions, when using Strang splitting the
state can drift significantly off of the smooth solution to the
coupled reactive hydrodynamics equations, as shown graphically in
\citet{astronum:2018} (using an earlier version of the present
algorithm).  
%A variation on Strang splitting called (re-)balanced
%splitting was developed in \citet{speth:2013}, but we do not explore
%that here. \MarginPar{keep this sentence? MK: Don't think it's needed.}

\subsection{Timestep Limiters and Retry Mechanism}

Since this method is based off of the CTU hydrodynamics scheme, it
benefits from the larger timestep that method can take (when done with
full corner coupling, the advective CFL condition is unity) as
compared to a method-of-lines approach; see \cite{ppmunsplit}.  In
addition to the standard CFL timestep limiter for explicit
hydrodynamics, timestep limiters based on
the energy generation or abundance changes over a timestep, such as
those introduced in \cite{prometheus}, are often used.  An energy limiter takes the
form of either:
\begin{equation}
\label{eq:dt:nuce_I}
\Delta t \le f_e\, \min_{i} \left\{\frac{e^n_{i}}{|\dot{S}_{i}^n|}\right\} \\
\end{equation}
or
\begin{equation}
\label{eq:dt:nuce_II}
\Delta t \le f_e\, \min_{i} \left\{ \frac{e^n_i}{|\int \dot{S}_i(t) dt|} \right\}
\end{equation}
where $i$ is the zone index and $f_e$ is a parameter used to control
the allowed change (e.g., $f_e = 0.5$ will limit the step such that
reactions can only increase the internal energy by 50\%).  The two
formulations differ in that Equation (\ref{eq:dt:nuce_I}) uses the
instantaneous evaluation of the energy generation rate at the start of
the timestep while Equation (\ref{eq:dt:nuce_II}) uses the integral of the
energy generation rate over the entire step (or last Strang half).
Usually these limiters are used in a reactive manner---the next step will see
a smaller timestep based on what happened during the current step, but the
current step is still accepted, even if it violated these conditions.  Our
goal is to avoid the need for these limiters by improving the coupling of hydrodynamics
and reactions.

\castro\ has the ability to reject a timestep if it detects a failure and retry
with smaller timesteps (subcycling to make up the original required timestep).
Among the conditions that can trigger this are density falling below zero during
advection, the ODE integration failing to converge in the implicit solve (due to
too many steps or the internal timestep falling below a minimum), or violation
of one of the timestep limiters during the step.  This means that the timestep
constraints are proactive instead of reactive.  The retry mechanism in \castro\
works with both the Strang and simplified-SDC integration scheme.  Retrying the
step is much more stringent that simply adjusting the next step, so we put a cap
on the number of retries allowed, aborting if we exceed the cap. 
%\MarginPar{describe in more detail? MK: While I do want us to evangelize the retry scheme, it %seems slightly out of place in this paper.  MZ: I do note later that the retry kicked in for %both Strang and SDC, so that's why I added it here.}

\subsection{Spectral Deferred Corrections}

The basic idea of spectral deferred corrections \citep{dutt:2000} is to divide each time step into sub-steps defined by high-order quadrature points (e.g.~Gauss-Lobatto) and iteratively correct the solution at each temporal node in order to reduce the integration error.
Each correction equation can be formulated using low-order integration techniques, such as forward or backward-Euler, with additional terms on each right-hand-side arising from integrating the residual over the sub-step from the previous iteration using high-order temporal integration.
Each iteration over all sub-steps increases the overall order of accuracy of the method by one, up to the underlying order of accuracy of the quadrature over the entire time step.
A number of SDC approaches have been developed
over the years targeting tighter multiphysics coupling, such as 
\cite{bourlioux2003high,SDC-old}.  In these approaches, different physical processes can be treated with different approaches (such as forward-Euler for advection and backward-Euler for reactions), and furthermore different physical processes can be sub-stepped between temporal nodes.

The approach that we implemented in \citet{castro_sdc} is based off a
multi-implicit, method of lines approach in \citet{bourlioux2003high}.
In contrast to those works, the focus of this paper is not higher-order integration, but tighter coupling
between advection and reactions in a second-order framework.
Thus, our simplified-SDC approach is based on the method described in \citet{SDC-old}
and a similar (but unpublished) implementation in the \maestroex\ simulation
code \citep{maestroex}.
The key point in this approach is that we are re-using the same numerical kernels 
for advection and reactions that are used in a Strang splitting approach,
but we are able to include, rather than exclude, the effects of the other 
physical processes in each correction equation solve.

\begin{figure}[t]
\plotone{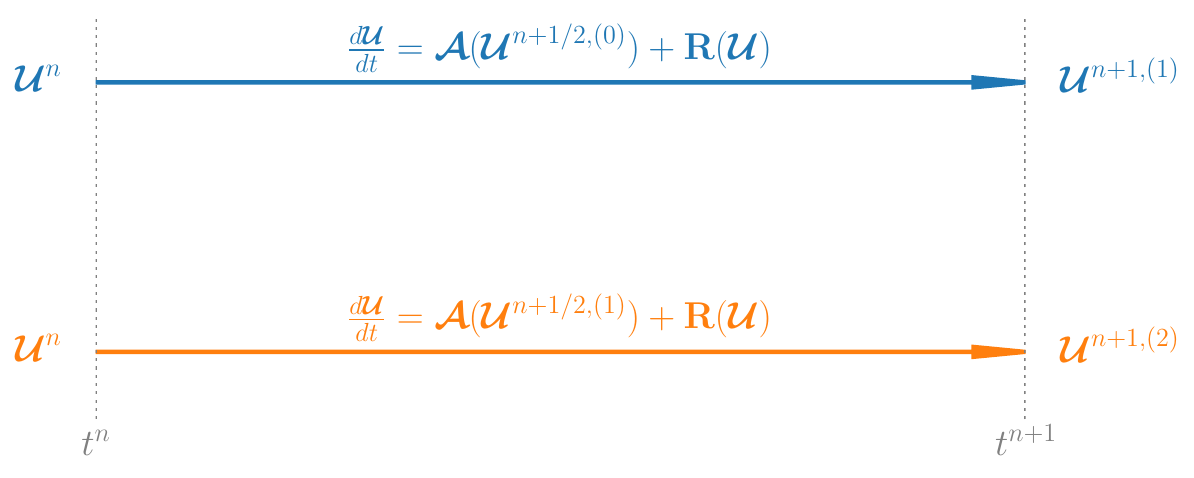}
\caption{\label{fig:sdc_flowchart} Flowchart of the simplified-SDC
  approach.  An iteratively-lagged advection term is used to update
  the system to the new time, with each iteration improving the
  coupling.}
\end{figure}

We begin with the general form of a multi-implicit SDC correction equation with two physical processes.
Using $(k)$ to denote the iterate, we have
\begin{eqnarray}
\Uc^{(k)}(t) = \Uc^n &+& \int_{t^n}^t \left [ \Advs{\Uc^{(k)}} + \Rb\left(\Uc^{(k)}\right) - \Advs{\Uc^{(k-1)}} - \Rb\left(\Uc^{(k-1)}\right) \right ] dt \nonumber\\
&+& \int_{t^n}^t \left [ \Advs{\Uc^{(k-1)}} + \Rb\left(\Uc^{(k-1)}\right) \right ] dt.
\end{eqnarray}
In the spirit of \cite{SDC-old}, we treat each advection term as piecewise constant over the time step,
and equal to a time-centered advection term evaluated with the Godunov integrator.
We note that this is a departure from the approach in \cite{bourlioux2003high} that uses a method of lines approach, where 
the advection terms are computed at each node, rather than the midpoint between nodes, using spatial extrapolation only.
Using $\boldsymbol{\mathcal{A}}(\Uc^{n+1/2})$ to denote the advection term, we have
\begin{eqnarray}
\Uc^{(k)}(t) = \Uc^n &+& \int_{t^n}^t \left [ \Advs{\Uc^{n+1/2,(k)}} + \Rb\left(\Uc^{(k)}\right) - \Advs{\Uc^{n+1/2,(k-1)}} - \Rb\left(\Uc^{(k-1)}\right) \right ] dt \nonumber\\
&+& \int_{t^n}^t \left [ \Advs{\Uc^{n+1/2,(k-1)}} + \Rb\left(\Uc^{(k-1)}\right) \right ] dt.
\end{eqnarray}
Also following \cite{SDC-old}, we differentiate this equation in time
\begin{eqnarray}
\frac{\partial\Uc^{(k)}}{\partial t} = && \left [ \Advs{\Uc^{n+1/2,(k)}} + \Rb\left(\Uc^{(k)}\right) - \Advs{\Uc^{n+1/2,(k-1)}} - \Rb\left(\Uc^{(k-1)}\right) \right]\nonumber\\
&+& \left [ \Advs{\Uc^{n+1/2,(k-1)}} + \Rb\left(\Uc^{(k-1)}\right) \right].
\end{eqnarray}
We combine terms, and arrive at a system of ODEs at each cell,
\begin{equation}
\frac{\partial\Uc^{(k)}}{\partial t} =  \left [ \Advs{\Uc^{n+1/2,(k)}} + \Rb\left(\Uc^{(k)}\right) \right].\label{eq:sdcupdate}
\end{equation}
We integrate this system from $t^n$ to $t^{n+1}$ using a modified version
of the VODE~\citep{vode} integrator, as described in Section \ref{sec:VODE}.
Thus, we are effectively using two temporal nodes ($t^n$ and $t^{n+1}$) 
and the overall integration is second-order accurate.
In order to properly couple reactions into the computation of the advection terms, 
we use an iteratively lagged reaction source term,
denoted $\Ic_q$ (this is basically an approximation of $\Rb(\qb)$).
The overall flow of the simplified-SDC algorithm is shown in Figure~\ref{fig:sdc_flowchart}.

%We start by considering the update in integral form:
%\begin{equation}
%\Uc^{n+1} = \Uc^n + \int_t^{t+\Delta t} \left [ \Advs{\Uc} + \Rb(\Uc) \right ] dt,
%\end{equation}
%In this approach, we approximate the advective term piecewise constant in time, using the value at the midpoint in time, to achieve second-order accuracy, giving us:
%\begin{equation}
%\label{eq:integral:simplesdc}
%\Uc^{n+1} = \Uc^n + \int_t^{t+\Delta t} \left \{ \Adv{\Uc^{n+1/2}} + \Rb(\Uc) \right \} dt
%\end{equation}
%The \castro\ version of the simplified-SDC algorithm differs from the previous versions due to the need to do some operations on the conserved variable state and some on the primitive variable state. 

\subsection{Time Advancement Scheme}
The basic time update algorithm proceeds as follows.
We begin with the state at $t^n$, denoted $\Uc^n$, and proceed as follows:

\begin{itemize}

%\item {\em Initialization}

%\label{sec:initialization}

%  \begin{itemize}
%  \item We need an approximation of how much the reactions alone
%    changed the primitive variable state over the timestep, which we
%    will denote $\Ic_q$ (this is basically an approximation of
%    $\Rb(\qb)$).

%    Since we do not have any information about the current
%    timestep in the first iteration, we use the value from the last
%    iteration of the previous timestep:
%    \begin{equation}
%      \Ic^{n+1/2,(0)}_{\qb} = \Ic^{n-1/2,(\kmax)}_{\qb}
%    \end{equation}

  %% \item Solve the Poisson problem for the initial gravitational potential:
  %%   \begin{equation}
  %%     \nabla^2 \Phi^n = 4\pi G \rho^n
  %%   \end{equation}

  %% \item Set the first guess at the new time potential as
  %%   $\Phi^{n+1,(0)} = \Phi^n$.

  %\end{itemize}

\item {\em Iterate}

  Iterate from $k = 1, \ldots, \kmax$.  For second-order accuracy,
  $\kmax = 2$ is sufficient; further SDC iterations will continue to decrease the splitting error,
  but not increase the formal order of accuracy of the method.  In addition to denoting the time-level
  with a superscript (like $n$ or $n+1$), we use a second superscript
  in parentheses to keep track of the iteration.  A single iteration, $k$,
  starts with $\Uc^n$ and results in the new time-level state for that
  iteration, $\Uc^{n+1,(k)}$.

  \begin{itemize}
  \item {\em Step 1: Create the advective update term, $\Adv{\Uc}^{n+1/2,(k)}$}

    \begin{itemize}
    \item Convert $\Uc \rightarrow \qb$.  This is an algebraic transformation,
      that utilizes the equation of state.

    \item Predict $\qb$ to the interfaces at $t^{n+1/2}$ using the CTU
      PPM method---this involves taking the cell-centered
        primitive state and tracking all of the information that can
        reach the interface over a timestep, following the procedure
        from \citet{ppmunsplit,millercolella:2002}.  The source
      terms, $\Sq$, used in the prediction are:
      \begin{equation}
        \Sc(\qb) = \Gb(\qb) + \Ic_\qb^{n+1/2,(k-1)}
      \end{equation}
      $\Ic_\qb^{n+1/2,(k-1)}$ is a numerical approximation for the effect of
      reactions from the previous iterate, which is computed at the end of each iterate 
      as described below.
      The use of this term contrasts with Strang-splitting, where no
      reactive source terms are included in the hydrodynamics update.
      If $k=1$, we instead use the value from the last iteration of the previous
      time step, $\Ic^{n-1/2,(\kmax)}_{\qb}$.
%%       Any hydrodynamic source terms are time-centered
%%       using the previous iteration:
%% \MarginPar{Formally both $t^n$ or $t^{n+1/2}$ sources give you 2nd-order since they are CTU predictor source terms.  I've never found this type of iteration for this term to amount to any benefit}
%%       \begin{equation}
%%         \Sqhydro^{n+1/2} = \frac{1}{2} \left ( \Sqhydro^n + \Sqhydro^{n+1,(k-1)} \right )
%%       \end{equation}

      In the unsplit CTU method \citep{ppmunsplit}, the interface
      states used for the final Riemann problem through the zone
      interface consist of a normal predictor and a transverse flux
      correction.  We can add the source terms either to the normal
      predictor (for example, doing characteristic tracing as
      described in \citealt{ppm}) or after all of the transverse flux
      corrections are made.  Both are second-order accurate.  For the
      gravitational sources, we do those in the normal predictor, consistent
      with the formulation in \citet{ppm}.
      However, for the reactive sources, we found that it is most
      reliable to add them at the end of the interface state
      construction, after the transverse flux corrections.  This is
      because we want to ensure that sum over species of $\Ic_\qb$ is
      zero, and characteristic tracing or the various flux corrections
      in general do not preserve this.  We enforce that the species interface
      states remain in $[0, 1]$ after adding the reactive source.

    \item Solve the Riemann problem at each interface to get a unique
      conserved state on each interface, $\Uc^{n+1/2,(k)}_{i+1/2}$.

    \item Construct the advective update terms, $\Advt{\Uc}^{n+1/2,(k)}_{i}$,
      first without the gravitational sources,
      \begin{equation}
        \Advt{\Uc}^{n+1/2,(k)}_{i} =
          - \frac{\Fb^\xv(\Uc^{n+1/2,(k)}_{i+1/2}) - \Fb^\xv(\Uc^{n+1/2,(k)}_{i-1/2})}{\Delta x}
      \end{equation}
      for the general case, and
      \begin{eqnarray}
        \left(\tilde{\mathcal{A}}_{\rho e}^{n+1/2,(k)}\right)_{i} &=&
          - \frac{\left(F_{\rho e}^{n+1/2,(k)}\right)_{i+1/2} - \left(F_{\rho e}^{n+1/2,(k)}\right)_{i-1/2}}{\Delta x} \nonumber \\
          &&- \left (\frac{p_{i+1/2}^{n+1/2,(k)} + p_{i-1/2}^{n+1/2,(k)}}{2} \right )
          \left(\frac{u_{i+1/2}^{n+1/2,(k)} - u_{i-1/2}^{n+1/2,(k)}}{\Delta x}\right)
      \end{eqnarray}
      for $(\rho e)$.

    Now the gravitational source terms, $\Gb(\Uc)$, are computed by first updating to the
    new state with advection and the old-time source term applied for the full $\Delta t$ as:
    \begin{equation}
      \Uc^{\star\star} = \Uc^n + \Delta t \Advt{\Uc}^{n+1/2,(k)} + \Delta t \Gb(\Uc^n)\label{eq:grav_pred}
    \end{equation}
    %% first updating the density to the new state with advection only as:
    %% \begin{equation}
    %%   \rho^{\star\star} = \rho^n + \Delta t \Advt{\rho}^{n+1/2,(k)}
    %% \end{equation}
    %% then constructing the momentum source term:
    %% \begin{equation}
    %%   {\bf S}_{\rho\Ub}^{n+1/2} = \frac{1}{2} (\rho^n \gb^n + \rho^{\star\star} \gb^{n+1,(k-1)})
    %% \end{equation}
    %% Then updating the momentum with only advection as:
    %% \begin{equation}
    %%   (\rho \Ub)^{\star\star} = (\rho \Ub)^n + \Delta t \Advt{\rho\Ub}^{n+1/2,(k)} + \Delta t {\bf S}_{\rho\Ub}^{n+1/2}
    %% \end{equation}
    %% and finally constructing the energy source:
    %% \begin{equation}
    %%   S_{\rho E}^{n+1/2} = \frac{1}{2} \left [ (\rho\Ub)^n \cdot \gb^n + (\rho\Ub)^{\star\star} \cdot \gb^{n+1,(k-1)}
    %%     \right ]
    %% \end{equation}
    We then evaluate the source terms with $\Uc^{\star\star}$ and
    correct the advective term so that we have a time-centered
    source.
    The final advective update term is then\footnote{For a source like gravity, this update can be done first for $\rho$ and then define the new momentum source using $\rho^{\star\star}$ and likewise for energy.}:
    \begin{equation}
      \Adv{\Uc}^{n+1/2,(k)}_{i} = \Advt{\Uc}^{n+1/2,(k)}_{i} +
      \frac{\Delta t}{2} \left [\Gb(\Uc^n) + \Gb(\Uc^{\star\star}) \right ]\label{eq:grav_corr}
    \end{equation}

    %% with
    %% \begin{equation}
    %%   \Gb^{n+1/2} = \left ( \begin{array}{c}
    %%                  0 \\ 0 \\ 0 \\
    %%                 {\bf S}_{\rho\Ub}^{n+1/2} \cdot \ex \\
    %%                 S_{\rho E}^{n+1/2} \end{array} \right )
    %% \end{equation}

    Formally, per \cite{SDC-old}, an alternative strategy would be to skip Equation (\ref{eq:grav_pred}) and instead
    perform only Equation (\ref{eq:grav_corr}) but with $\Gb(\Uc^{n+1,(k-1)})$ replacing $\Gb(\Uc^{\star\star})$.
    In the future we will explore the efficacy of both approaches.
    %\MarginPar{AJN: rewrote this; ok?  If the alternative works it's actually much better since %fewer gravity solves are required.}
    %We note that this procedure works because the source terms here do not depend on the outcome of the reactions.   In \cite{SDC-old}, instead of using $\Gb(\Uc^{\star\star})$, $\Gb(\Uc^{n+1,(k-1)})$ is used.  For the conservative gravity formulation in \cite{wdmergerI}, we prefer the former formulation since the source terms are computed using the result of the Riemann solver.\MarginPar{MK: I think this should be written to say that we are *assuming* the source terms do not depend on the outcome of the reactions, or clarify that it is in fact true specifically in Castro that we only use source terms where this is true. But actually, wouldn't thermal diffusion depend on the outcome of reactions? MZ: I am not sure anymore what the intent was here.  I think we should just remove this sentence.  I don't think that it affects diffusion.  I think it meant that $\omegadot$ does not explicitly appear..}

    \end{itemize}

  \item {\em Step 2: Integrate Using VODE}
  %\MarginPar{Need to change this.  The point of this paper is we are NOT using method of lines (that was the older castro sdc paper).  This simplified approach is NOT method of lines.}

    We update the state by integrating equation
    (\ref{eq:sdcupdate}) over the full time step.  Since we are approximating
    $\boldsymbol{\mathcal{A}}(\Uc)$ as piecewise constant in time, we can use an
    ODE integrator to integrate this, just as we do with the reaction
    system in Strang splitting.  The difference here is that we are
    integrating the conserved variables and the state sees the effect
    of advection and gravity as we integrate the reactions.  So rather than use
    $d\Uc/dt = \Rb(\Uc)$ as in the Strang case, the ODE form we use is
    \begin{equation}
      \odt{\Uc} = \Adv{\Uc}^{n+1/2,(k)} + \Rb(\Uc)
    \end{equation}
    The details of the VODE~\citep{vode} integrator we use are
    described in Section \ref{sec:VODE}.
    
    Looking at the form of Equation (\ref{eq:cons_sources}), we see that only
    the species and energies have non-zero terms in $\Rb$.  The total and
    internal energies both provide the same information, and
    integrating both overconstrains the system, so we just integrate
    the internal energy.  We define the subset of variables that are
    directly integrated as:
    \begin{equation}
      \Uc^\prime = \left ( \begin{array}{c} \rho X_k \\ \rho e \end{array} \right )
    \end{equation}
    and we integrate
    \begin{equation}
      \odt{\Uc^\prime} = \Adv{\Uc^\prime}^{n+1/2,(k)} + \Rb(\Uc^\prime)
    \end{equation}
    This integration begins with $\Uc^{\prime,n}$ and results in $\Uc^{\prime,n+1,(k)}$.

    We will need the density at times during the integration, which we construct as:
    \begin{equation}
      \rho(t) = \rho^n + \mathcal{A}_\rho^{n+1/2,(k)} \, (t - t^n)
    \end{equation}
    As we are integrating this system we need to get the temperature,
    $T$, for the rate evaluations.  We obtain this directly from
    internal energy, composition, and density using the equation of
    state.

    Our integrator also needs the Jacobian of the system, in terms of
    $\Uc^\prime$.  This is different than the form of the Jacobian
    usually used in reaction networks (we depend on $e$ instead of
    $T$).  We describe the form of the Jacobian in Appendix
    \ref{sec:app:jacobian}.

    At the end of the integration, we can do the conservative update of momentum
    and energy.  Momentum is straightforward, since there are no reactive sources:
    \begin{equation}
      (\rho u)^{n+1,(k)} = (\rho u)^n + \Delta t \mathcal{A}_{\rho u}^{n+1/2,(k)}
    \end{equation}
    For total energy, we first need to isolate the reactive source for energy:
    \begin{equation}
      (\rho \Sdot)^{n+1/2,(k)} = \frac{(\rho e)^{n+1,(k)} - (\rho e)^n}{\Delta t} - \mathcal{A}_{\rho e}^{n+1/2,(k)}
    \end{equation}
    Then the update is
    \begin{equation}
      (\rho E)^{n+1,(k)} = (\rho E)^n + \Delta t \mathcal{A}_{\rho E}^{n+1/2,(k)} + \Delta t (\rho \Sdot)^{n+1/2,(k)}
    \end{equation}

  \item {\em Step 3: Compute the Reactive Source Terms, $\Ic^{n+1/2,(k)}$}

    We now seek the $\Ic$'s that capture the effect of just the
    reaction sources on the state variables for the next iteration.
    For the conserved quantities, these would simply be:
    \begin{equation}
      \Ic^{n+1/2,(k)}_{\Uc} = \frac{\Uc^{n+1,(k)} - \Uc^n}{\Delta t} - \Adv{\Uc}^{n+1/2,(k)}
    \end{equation}
    However, for our primitive variables, which are used in the
    interface state prediction, we need to construct the required source terms more
    carefully.  We want:
    \begin{equation}
      \Ic^{n+1/2,(k)}_{\qb} = \frac{\qb^{n+1,(k)} - \qb^n}{\Delta t} - \Adv{\qb}^{n+1/2,(k)}\label{eq:Iq_prim}
    \end{equation}
    but we need the advective update for $\qb$, which we have not
    constructed.  We note that $\Ic^{n+1/2,(k)}_\qb$ is an approximation to the integral of
    Equation (\ref{eq:prim_sources}) over the timestep.  Additionally, we cannot simply use the equation of
    state on $\Ic^{n+1/2,(k)}_{\Uc}$ since this is a time-derivative and
    does not represent a well-defined state in itself.  Instead, we
    derive $\Ic^{n+1/2,(k)}_{\qb}$ via a multi-step process.  We first find
    the conservative state as if it were updated only with advection:
    \begin{equation}
      \Uc^\star = \Uc^n + \Delta t \Adv{\Uc}^{n+1/2,(k)}
    \end{equation}
    and then construct the corresponding primitive variable state via an algebraic transform,
    $\Uc^\star \rightarrow \qb^\star$.
    This allows us to define the advective update for $\qb$ as:
    \begin{equation}
      \Adv{\qb}^{n+1/2,(k)} = \frac{\qb^\star - \qb^n}{\Delta t}\label{eq:Aq_prim}
    \end{equation}
    Defining the primitive state corresponding to the fully-updated
    conserved state via an algebraic transform, $\Uc^{n+1,(k)}
    \rightarrow \qb^{n+1,(k)}$, we can construct $\Ic^{n+1/2,(k)}_{\qb}$ by
    combining Equations (\ref{eq:Iq_prim}) and (\ref{eq:Aq_prim}),
    \begin{equation}
      \Ic^{n+1/2,(k)}_{\qb} = \frac{\qb^{n+1,(k)} - \qb^\star}{\Delta t}.
    \end{equation}

  %% \item {\em Solve for the New Gravitational Potential.}

  %%   We solve
  %%   \begin{equation}
  %%     \nabla^2 \Phi^{n+1,(k)} = 4\pi G \rho^{n+1,(k)}
  %%   \end{equation}

  \end{itemize}

\end{itemize}

This completes the update of a single iteration.

\subsection{Stiff ODE Integrator}\label{sec:VODE}

We use a modified version of the VODE~\citep{vode} integrator,
designed for stiff ODEs.  Our version has been ported to modern C++ and designed to run on GPUs.  We have also added
checks to the timestep rejection logic that help prevent VODE from
wandering too far off the solution\footnote{This modified version of
VODE is available in our Microphysics repo:
\url{https://github.com/AMReX-Astro/Microphysics}.}.  We have found empirically that these checks improve the performance of VODE.

VODE allows us to specify both relative and absolute tolerances
for the species and energy during the integration.  They are combined
into a weight that VODE uses to assess convergence of the step:
\begin{equation}
w_i = \epsilon_\mathrm{rel} |y_i| + \epsilon_\mathrm{abs}
\end{equation}
for integration variable $y_i$.  We will denote the species tolerances
as $\rtolspec$ and $\atolspec$ and the energy tolerances as $\rtolener$ and
$\atolener$.

For Strang splitting, we require that
an individual mass fraction not change too much over a step:
\begin{equation}
\label{eq:vode_species_reject}
\begin{array}{c}|X_k^{m+1}| < f |X_k^m| \\ |X_k^{m+1}| > \frac{1}{f} |X_k^m| \end{array} ~  \mbox{if}~ |X_k^m| > \eta \atolspec ~\mbox{and}~ |X_k^{m+1}| > \eta \atolspec
\end{equation}
Here, $m$ is the current VODE solution and $m+1$ is the potential
new-time solution.  The parameter $f$ is the factor a mass fraction is
allowed to change per VODE step.  We use $f = 4$ for the results here.
The parameter $\eta$ allows us to only use this condition for species
with mass fractions above a threshold.  We use $\eta = 1$ by default,
but $\eta = f$ sometimes helps prevent VODE from worrying about
species that dip below the threshold during integration, since VODE
uses a norm over all of the integrated variables to compute the total
error.  If these conditions are violated, then VODE will reject the
timestep and retry with a smaller step.  We also require that the mass
fractions are contained in $[0, 1]$ to a tolerance of $10^{-2}$.
For SDC, we enforce these same constraints, but for
Equation (\ref{eq:vode_species_reject}), we only use the change from $m$ to $m+1$ due
to reactions, by subtracting off the advective contribution over that substep.
Finally, we enforce that:
\begin{align}
\rho^{m+1} &> 0 \\
(\rho e)^{m+1} &> 0
\end{align}

Some of the rates in \citet{reaclib} increase by hundreds of orders of
magnitude at low temperature, presumably because the interpolant fit was done at
higher temperatures.  Since VODE can take exploratory right-hand side evaluations
outside of the nominal time range, it can encounter these poorly-behaved rates
and it can cause the integration to fail (by generating {\tt inf}s).  To prevent
this behavior, we set a lower temperature limit of the reaction rates, below
which we zero them out.  For the simulations in section~\ref{sec:dd}, we choose a cutoff of
$5\times 10^7~\mathrm{K}$.

For both Strang splitting and simplified-SDC, we use a numerical Jacobian that VODE
computes internally (following the algorithm in \citealt{lsode}).  For
completeness, we derive the analytic form of the Jacobian in Appendix
\ref{sec:app:jacobian} for the simplified-SDC method. VODE employs
Jacobian-caching, so the Jacobian does not need to be re-evaluated each time it
is used.

\section{Tests}

\subsection{Reacting convergence test problem}

\citet{castro_sdc} introduced a test problem for measuring convergence
of a reacting hydrodynamic algorithm.  This is based on the acoustic
pulse problem from \citet{mccorquodalecolella}, and was also used in
\citet{strang_rnaas}.  A periodic domain with a uniform entropy is
initialized with a pressure profile with a small perturbation.  This
drives a low amplitude acoustic wave radially outward from the center.
A simple reaction network with the triple-alpha and
$\isotm{C}{12}(\alpha,\gamma)\isotm{O}{16}$ rates releases energy.  By
running at different spatial resolutions with a fixed CFL number so that the timestep scales
with resolution, we can compute the convergence rate in space and time.
In order to demonstrate second-order convergence, the problem must be smooth (so
the slope limiters don't kick in severely), which means that the
energy release cannot be so vigorous as to drive shocks.

% EOS: /home/zingale/development/Microphysics/EOS/helmholtz
% NETWORK: /home/zingale/development/Microphysics/networks/triple_alpha_plus_cago

% Castro       git describe: 21.12-9-gd7a27ccd5
% AMReX        git describe: 21.12-64-gfde72340e
% Microphysics git describe: 21.12-12-g9595a34f

\begin{deluxetable}{lllllll}
\tablecaption{\label{table:react_converge_strang} Convergence ($L_1$ norm) for the reacting convergence problem using Strang splitting.}
\tablehead{\colhead{field} & \colhead{$\epsilon_{64 \rightarrow 128}$} &
           \colhead{rate} & \colhead{$\epsilon_{128\rightarrow 256}$} &
           \colhead{rate} & \colhead{$\epsilon_{256\rightarrow 512}$}}
\startdata
 $\rho$                      & $2.794 \times 10^{18}$  & 2.044  & $6.777 \times 10^{17}$  & 2.554  & $1.154 \times 10^{17}$  \\
 $\rho u$                    & $6.796 \times 10^{26}$  & 2.448  & $1.245 \times 10^{26}$  & 2.889  & $1.681 \times 10^{25}$  \\
 $\rho v$                    & $6.796 \times 10^{26}$  & 2.448  & $1.245 \times 10^{26}$  & 2.889  & $1.681 \times 10^{25}$  \\
 $\rho E$                    & $2.451 \times 10^{35}$  & 2.351  & $4.803 \times 10^{34}$  & 2.742  & $7.179 \times 10^{33}$  \\
 $\rho e$                    & $2.261 \times 10^{35}$  & 2.320  & $4.526 \times 10^{34}$  & 2.821  & $6.403 \times 10^{33}$  \\
 $T$                         & $2.237 \times 10^{21}$  & 1.691  & $6.927 \times 10^{20}$  & 2.482  & $1.240 \times 10^{20}$  \\
 $\rho X(\isotm{He}{4})$     & $2.878 \times 10^{18}$  & 2.020  & $7.096 \times 10^{17}$  & 2.529  & $1.229 \times 10^{17}$  \\
 $\rho X(\isotm{C}{12})$     & $1.698 \times 10^{17}$  & 1.950  & $4.393 \times 10^{16}$  & 2.232  & $9.353 \times 10^{15}$  \\
 $\rho X(\isotm{O}{16})$     & $1.687 \times 10^{14}$  & 1.660  & $5.338 \times 10^{13}$  & 1.957  & $1.375 \times 10^{13}$  \\
 $\rho X(\isotm{Fe}{56})$    & $2.794 \times 10^{8}$   & 2.044  & $6.777 \times 10^{7}$   & 2.554  & $1.154 \times 10^{7}$   \\
\enddata
\end{deluxetable}

\begin{deluxetable}{lllllll}
\tablecaption{\label{table:react_converge_sdc} Convergence ($L_1$ norm) for the reacting convergence problem using SDC integration (2 iterations).}
\tablehead{\colhead{field} & \colhead{$\epsilon_{64 \rightarrow 128}$} &
           \colhead{rate} & \colhead{$\epsilon_{128\rightarrow 256}$} &
           \colhead{rate} & \colhead{$\epsilon_{256\rightarrow 512}$}}
\startdata
 $\rho$                      & $2.784 \times 10^{18}$  & 2.048  & $6.734 \times 10^{17}$  & 2.558  & $1.143 \times 10^{17}$  \\
 $\rho u$                    & $6.779 \times 10^{26}$  & 2.447  & $1.243 \times 10^{26}$  & 2.895  & $1.671 \times 10^{25}$  \\
 $\rho v$                    & $6.779 \times 10^{26}$  & 2.447  & $1.243 \times 10^{26}$  & 2.895  & $1.671 \times 10^{25}$  \\
 $\rho E$                    & $2.450 \times 10^{35}$  & 2.353  & $4.795 \times 10^{34}$  & 2.737  & $7.193 \times 10^{33}$  \\
 $\rho e$                    & $2.256 \times 10^{35}$  & 2.320  & $4.518 \times 10^{34}$  & 2.817  & $6.414 \times 10^{33}$  \\
 $T$                         & $2.232 \times 10^{21}$  & 1.695  & $6.893 \times 10^{20}$  & 2.484  & $1.233 \times 10^{20}$  \\
 $\rho X(\isotm{He}{4})$     & $2.866 \times 10^{18}$  & 2.024  & $7.049 \times 10^{17}$  & 2.534  & $1.217 \times 10^{17}$  \\
 $\rho X(\isotm{C}{12})$     & $1.695 \times 10^{17}$  & 1.959  & $4.357 \times 10^{16}$  & 2.236  & $9.250 \times 10^{15}$  \\
 $\rho X(\isotm{O}{16})$     & $1.681 \times 10^{14}$  & 1.662  & $5.313 \times 10^{13}$  & 1.963  & $1.363 \times 10^{13}$  \\
 $\rho X(\isotm{Fe}{56})$    & $2.784 \times 10^{8}$   & 2.048  & $6.734 \times 10^{7}$   & 2.558  & $1.143 \times 10^{7}$   \\
\enddata
\end{deluxetable}

Tables \ref{table:react_converge_strang} and
\ref{table:react_converge_sdc} show the Strang splitting and
simplified-SDC convergence respectively.  We ran each integration
method on a domain with $64^2$, $128^2$, $256^2$, $512^2$ zones.  An
error is defined between successive resolutions by coarsening the
higher resolution run and taking the L1 norm of the zone-by-zone
difference of the two simulations---for example,
$\epsilon_{64\rightarrow 128}$ is the error between the $64^2$ and
$128^2$ simulations. The convergence rate from
    $\epsilon_{64\rightarrow 128}$ to $\epsilon_{128\rightarrow 256}$
    is computed as $\log_2 \epsilon_{64\rightarrow
      128}/\epsilon_{128\rightarrow 256}$ and is given in the column
    between those errors, and likewise for the higher-resolution
    case.  A rate of 2.0 indicates second order convergence.  We see
  that both methods exhibit convergence at roughly second-order
    and agree well.  This demonstrates that both methods are working
  as designed.

\subsection{Sub-Chandra Double Detonation}
\label{sec:dd}

%\MarginPar{AJN: do you have any wall-clock, or RHS-count, or Nonaka-plot performance metrics?  SDC should not only give better scientific results, but should be faster as well...  I address this with the new plots in the last section now.}
We next model a double detonation Type Ia supernova.  In this model, a
detonation begins in the accreted helium layer on a sub-Chandrasekhar mass white
dwarf and that detonation can either drive a compression wave into the core
igniting a second detonation in the carbon / oxygen core, or burn through the
interface at the base of the accreted layer producing an inward propagating
carbon detonation (see, e.g., \citealt{fink:2007}).  Our goal here is not to
explore the feasibility of the model or understand whether a detonation
propagating through the He-C interface is physical, but rather to look at the
coupling of the hydrodynamics and reactions.  We want to drive vigorous burning
and directly compare Strang splitting and simplified-SDC time integration on this problem.
For that reason, we explore only a single model and start off with a rather
large temperature perturbation in the He layer.

We use a network with 28 nuclei and 107 rates that captures He burning with
links for $(\alpha,p)(p,\gamma)$ reactions and the rates involving \isot{N}{14}
to bypass $\isotm{C}{12}(\alpha,\gamma)\isotm{O}{16}$ as described in
\cite{shenbildsten}.  The rates are taken from the ReacLib library
\citep{reaclib} and the network is written by \pynucastro~\citep{pynucastro}
directly in the C++ format that our code requires.  The details of the network
are given in Appendix \ref{sec:app:reactionnet}. This moderate-sized network
with many reaction pathways (forward and reverse) make it a good test for the
coupling between hydro and reactions.

\begin{figure}[t]
\centering
\plotone{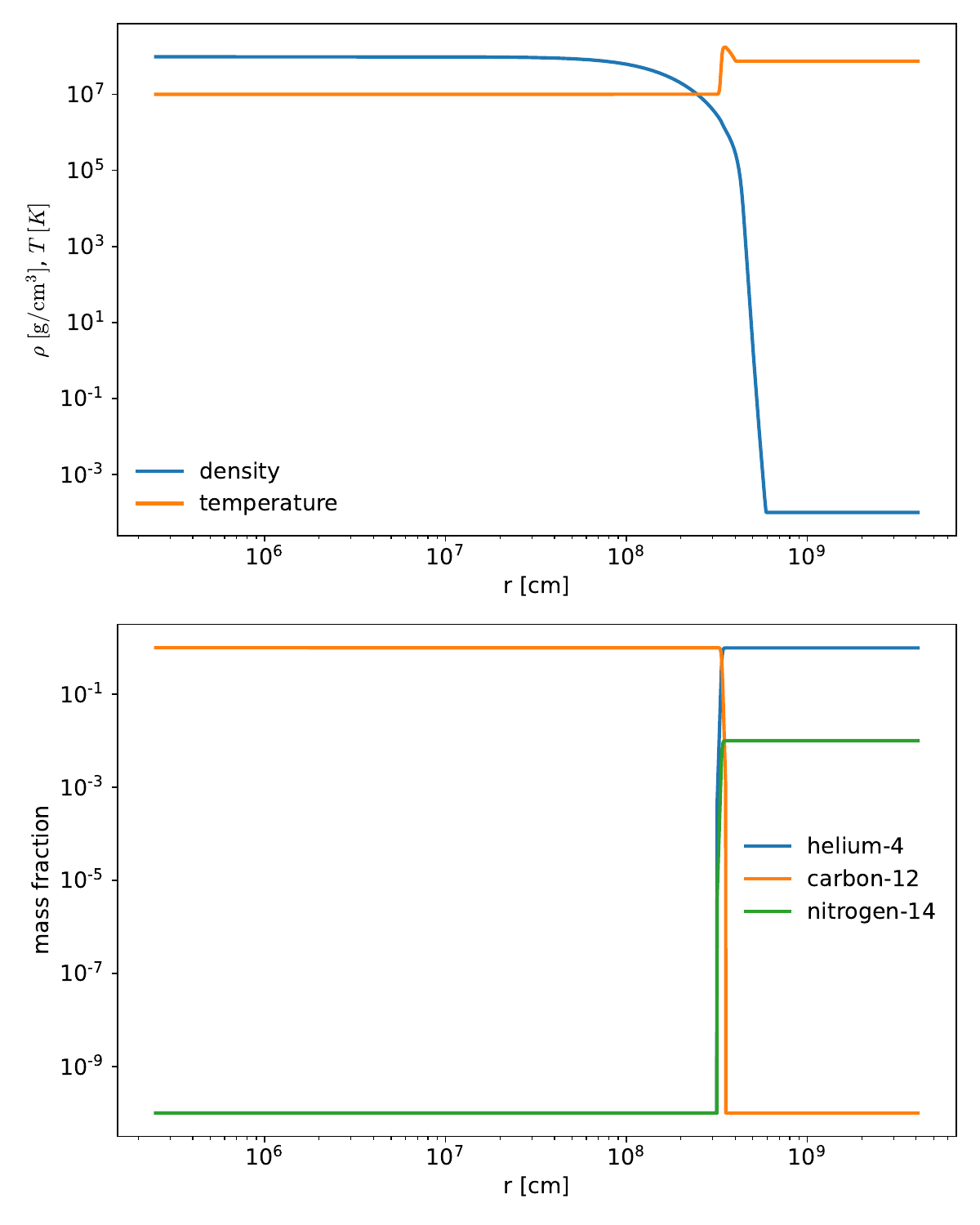}
\caption{\label{fig:subch_initial_model} Initial model for the double detonation test problem.}
\end{figure}

The initial model is generated following the procedure described in
\citep{subchandra}.  We use a $1.1~M_\odot$ carbon white dwarf with a
$0.05~M_\odot$ He layer.  We choose a pure \isot{C}{12} white dwarf to make the reactions more vigorous, consistent with our focus on the coupling of hydrodynamics and reactions.  We seed the He layer with 1\% (by mass) of
\isot{N}{14}.  The white dwarf is isothermal and the temperature ramps
up at the base of the He layer and is then isentropic until the
surface.  Figure~\ref{fig:subch_initial_model} shows the structure of
the initial model.  We use 2D axisymmetric coordinates with a domain
size of $1.024\times 10^9~\mathrm{cm}$ by $2.048\times
10^9~\mathrm{cm}$, a base grid of $256 \times 512$ zones and a single
level of refinement (jumping by $2\times$), giving a fine-grid
resolution of $20~\mathrm{km}$.  Gravity is modeled as a monopole.
Our choice of initial perturbation has the detonation propagate inward
into the carbon white dwarf.  We run the simulation for 0.1 s---this
is enough time for the carbon detonation to be well developed.

An initial perturbation is placed on the symmetry axis of the form:
\begin{equation}
  T = T_0 \left [ 1 + X(\isotm{He}{4}) f (1 + \tanh(2 - r_1)) \right ]
\end{equation}
where
\begin{equation}
  r_1 = \left [ x^2 + (y - r_0)^2 \right ]^{1/2} / \lambda
\end{equation}
and
\begin{equation}
  r_0 = r_\mathrm{pert} + r_\mathrm{base}
\end{equation}
Here, $r_\mathrm{base}$ is the radius at which the helium layer begins
and $r_\mathrm{pert}$ is the distance above the base to put the perturbation.  We
choose $r_\mathrm{pert} = 100~\mathrm{km}$.  The temperature is perturbed above
the initial model value, denoted as $T_0$ here.  The amplitude of the perturbation
is $f = 12$ and the scale of the perturbation is $\lambda = 200~\mathrm{km}$.

We do a suite of runs with both Strang splitting and the simplified-SDC integration, using
CFL numbers of 0.1, 0.2, and 0.4 and reaction network tolerances of $\atolspec =
\rtolspec = 10^{-5}$ and $\atolener = \rtolener = 10^{-5}$.  We also run the CFL
= 0.2 case with a tighter tolerance, using  $\atolspec = 10^{-8}$.  For the step
rejection part of VODE, we use $f = 4$ and $\eta = 4$.  The runs are labeled with
the prefix {\tt sdc\_} or {\tt strang\_} to denote the integrator, and also include
the CFL number in the name.  Finally, the runs with the tighter tolerances have
the suffix {\tt \_tol1.e-8}.  We note that both the Strang and the SDC runs encounter
VODE integration failures occasionally and utilize the \castro\ step retry functionality.

\begin{figure}
\centering
\plotone{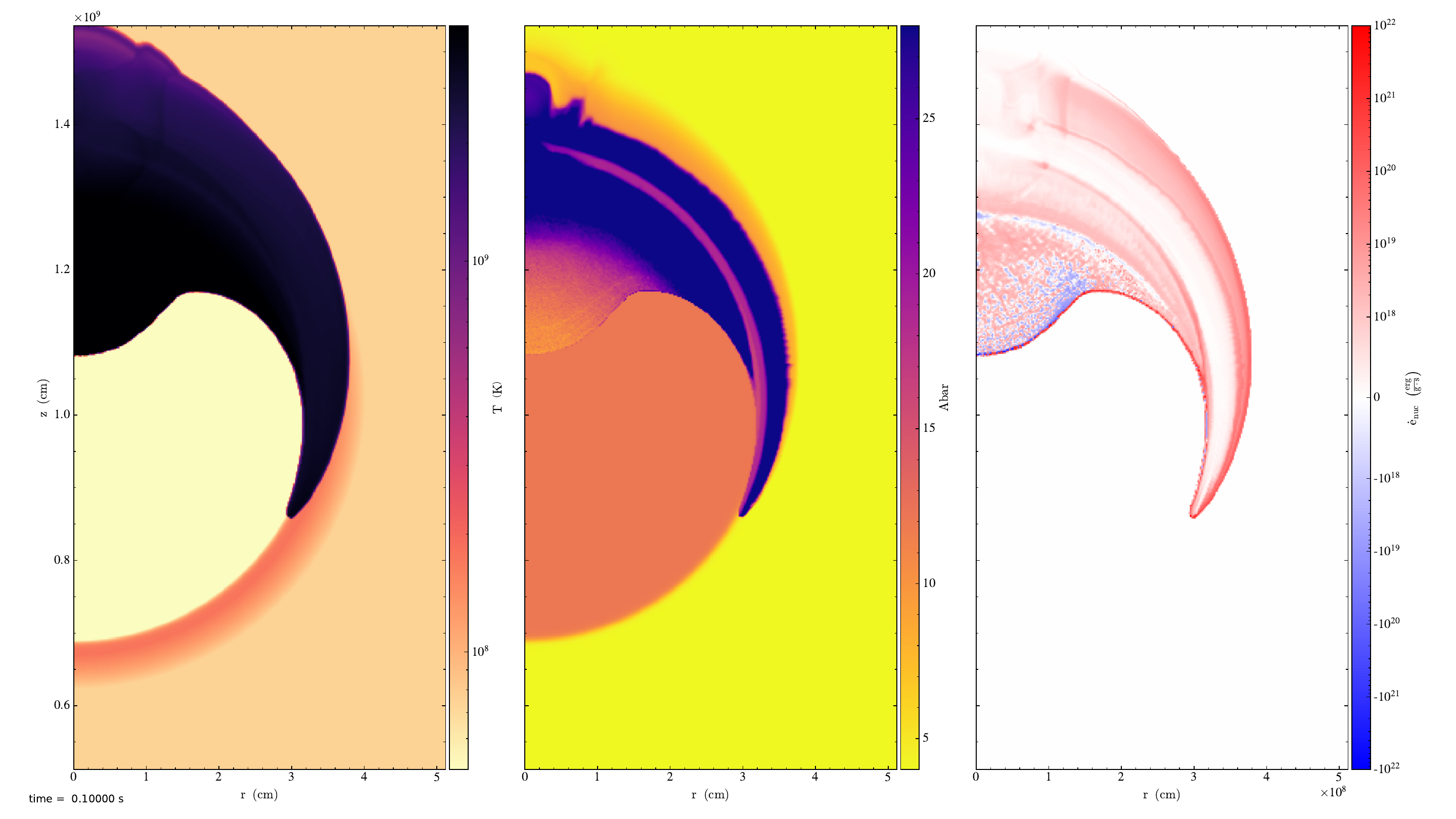}
\caption{\label{fig:subch_strang_cfl02} Temperature, mean molecular weight, and
energy generation rate for the Strang CFL = 0.2 run with normal tolerances ({\tt
strang\_subch2\_cfl0.2}) at 0.1 s.}
\end{figure}

\begin{figure}
\centering
\plotone{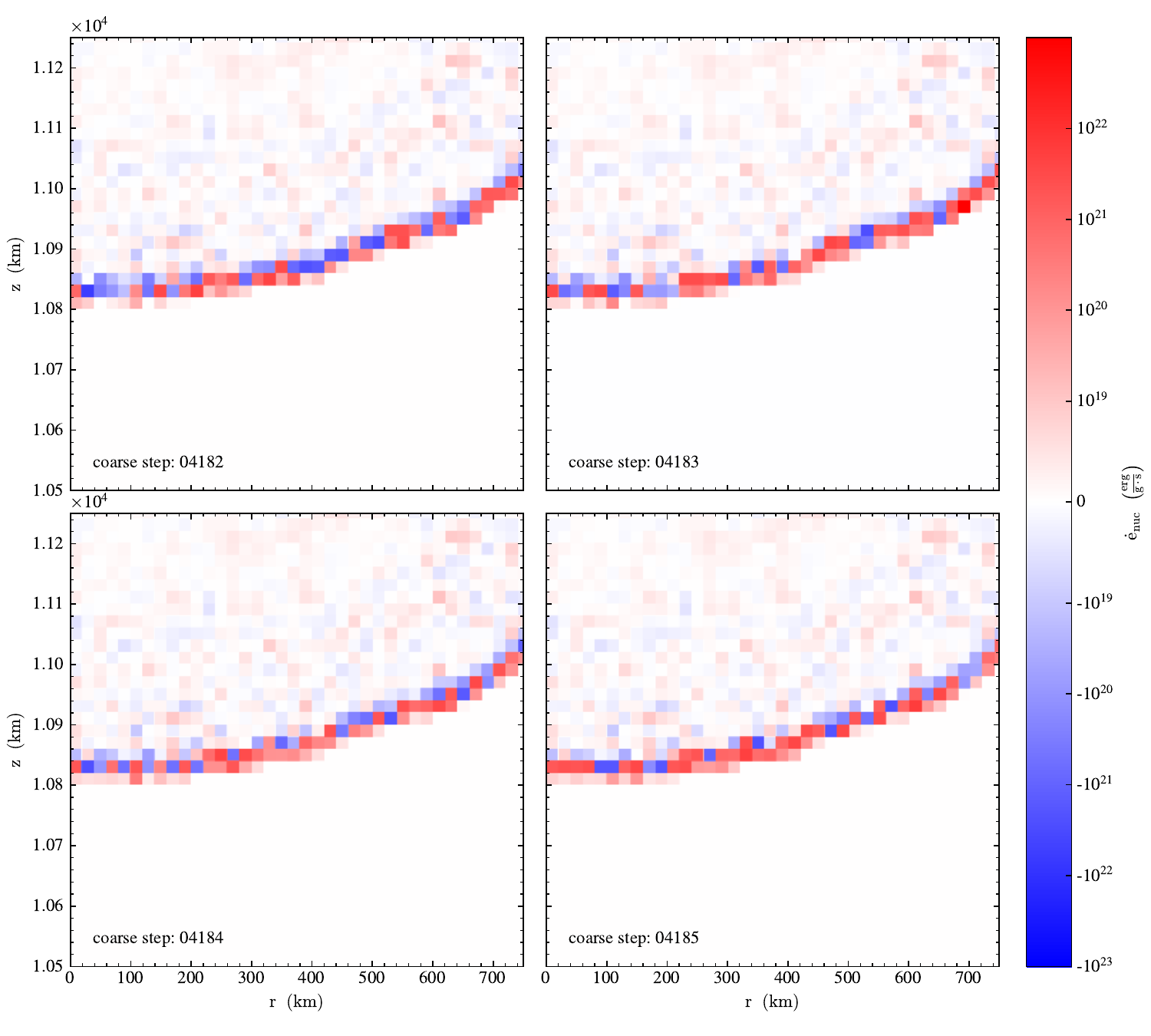}
\caption{\label{fig:strang_steps} Energy generation rate for the
  Strang CFL = 0.2 run with normal tolerances ({\tt
    strang\_subch2\_cfl0.2}) shown at 4 consecutive timesteps around
  $t = 0.1~\mathrm{s}$ zoomed in at the inward propagating detonation.}
\end{figure}

Figure \ref{fig:subch_strang_cfl02} shows the temperature, mean
molecular weight, and nuclear energy generation rate for the Strang splitting
CFL = 0.2 simulation ({\tt strang\_subch2\_cfl0.2}).  At this point
in the evolution, the helium detonation has wrapped more than halfway
across the surface and the ingoing carbon detonation is approaching the
center of the star.  We see that the
mean molecular weight appears mottled behind the inward propagating
detonation in the white dwarf.  Comparing to the energy generation
rate, we see that region is characterized by large energy releases of
alternating signs checkerboarding thoughout the region.  This is
characteristic of the nucleosynthesis struggling to come into
equilibrium at high temperatures when the reverse rates can be
important.  To highlight this, Figure~\ref{fig:strang_steps} shows the nuclear
energy generation rate in a highly-zoomed region around the detonation over four
consecutive coarse timesteps.  We see that on the detonation front itself, the
sign of the energy generation rate changes from one step to the next in many zones.

\begin{figure}
\centering
\plotone{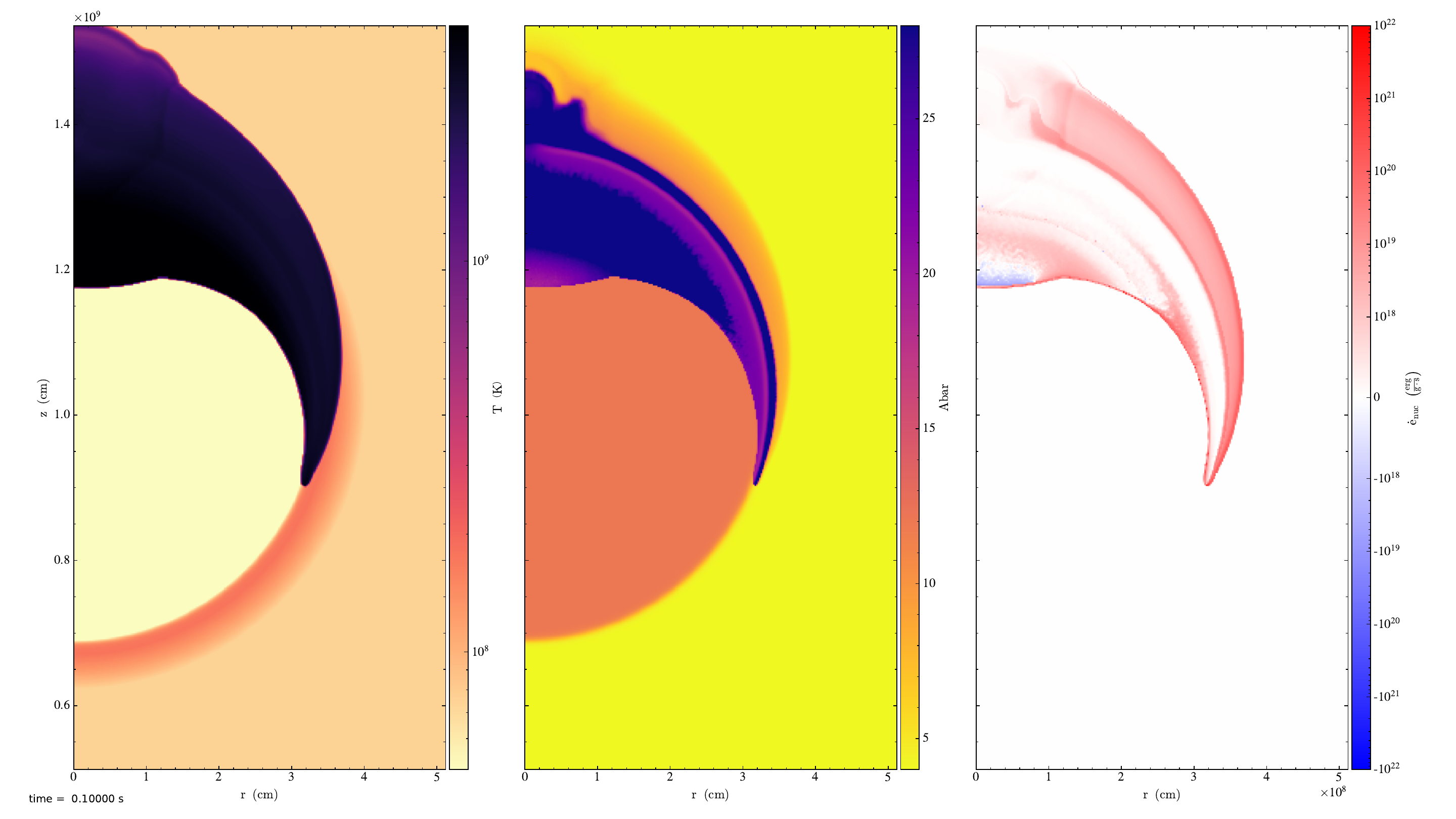}
\caption{\label{fig:subch_sdc_cfl02} Temperature, mean molecular weight, and
energy generation rate for the simplified-SDC CFL = 0.2 run with normal tolerances ({\tt
sdc\_subch2\_cfl0.2}) at 0.1 s.}
\end{figure}

We next look at the simplified-SDC version of this same case, {\tt
sdc\_subch2\_cfl0.2}.  Figure~\ref{fig:subch_sdc_cfl02} shows the temperature,
mean molecular weight, and energy generation rate. Compared to
Figure~\ref{fig:subch_strang_cfl02}, we see that the composition and energy
generation rates are smooth, without any of the checkerboarding that plagued the
Strang simulation.  This suggests that the simplified-SDC algorithm more accurately
attains the correct equilibrium in these conditions. More
strikingly, we see that the inward propagating carbon detonation has not
progressed as far into the white dwarf at this time.  The two integration
methods are giving very different results here.

\begin{figure}
\plotone{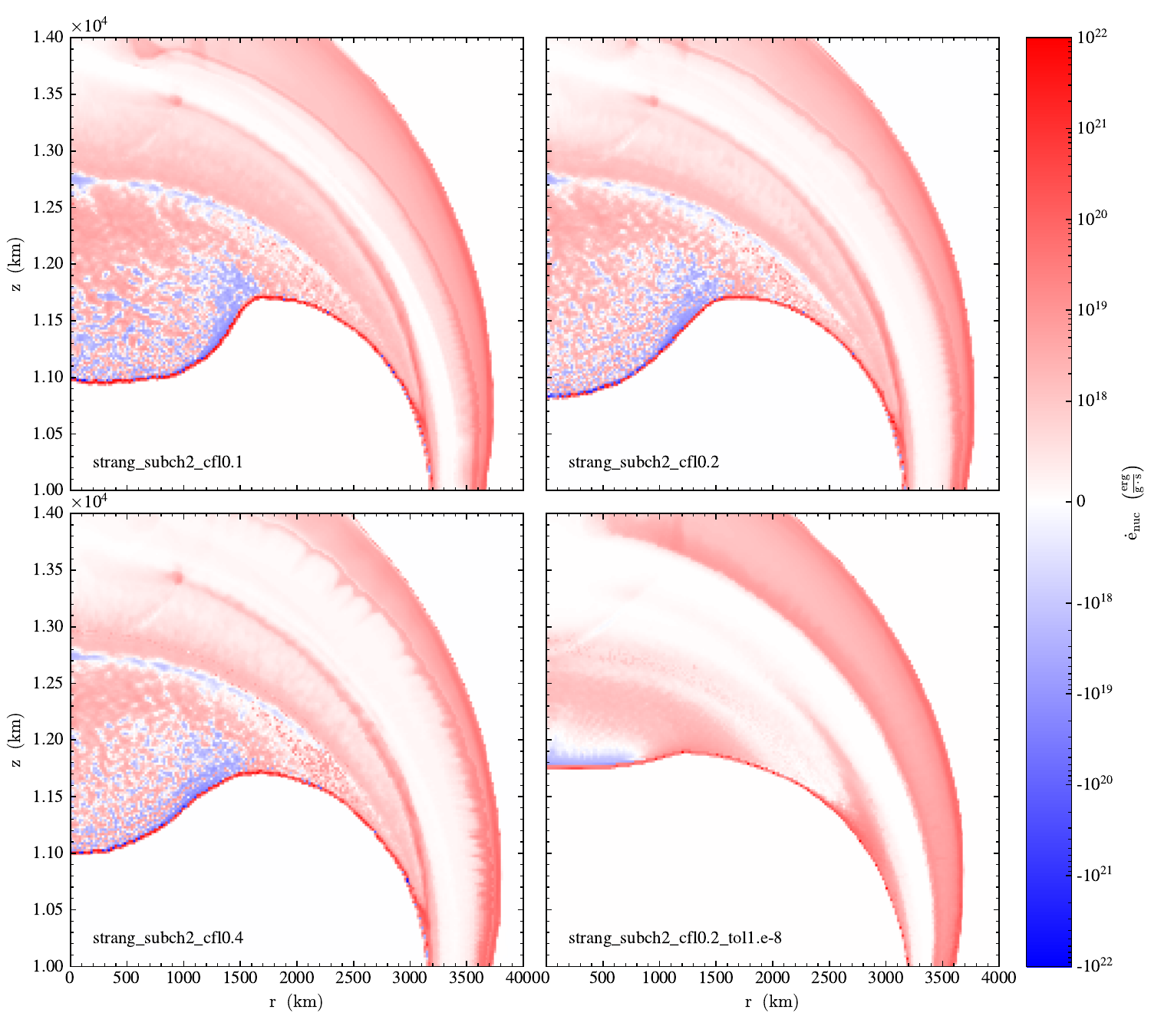}
\caption{\label{fig:strang_enuc_summary} Comparison of Strang runs showing
the nuclear energy generation rate behind the inward propagating
shock.}
\end{figure}

\begin{figure}
\plotone{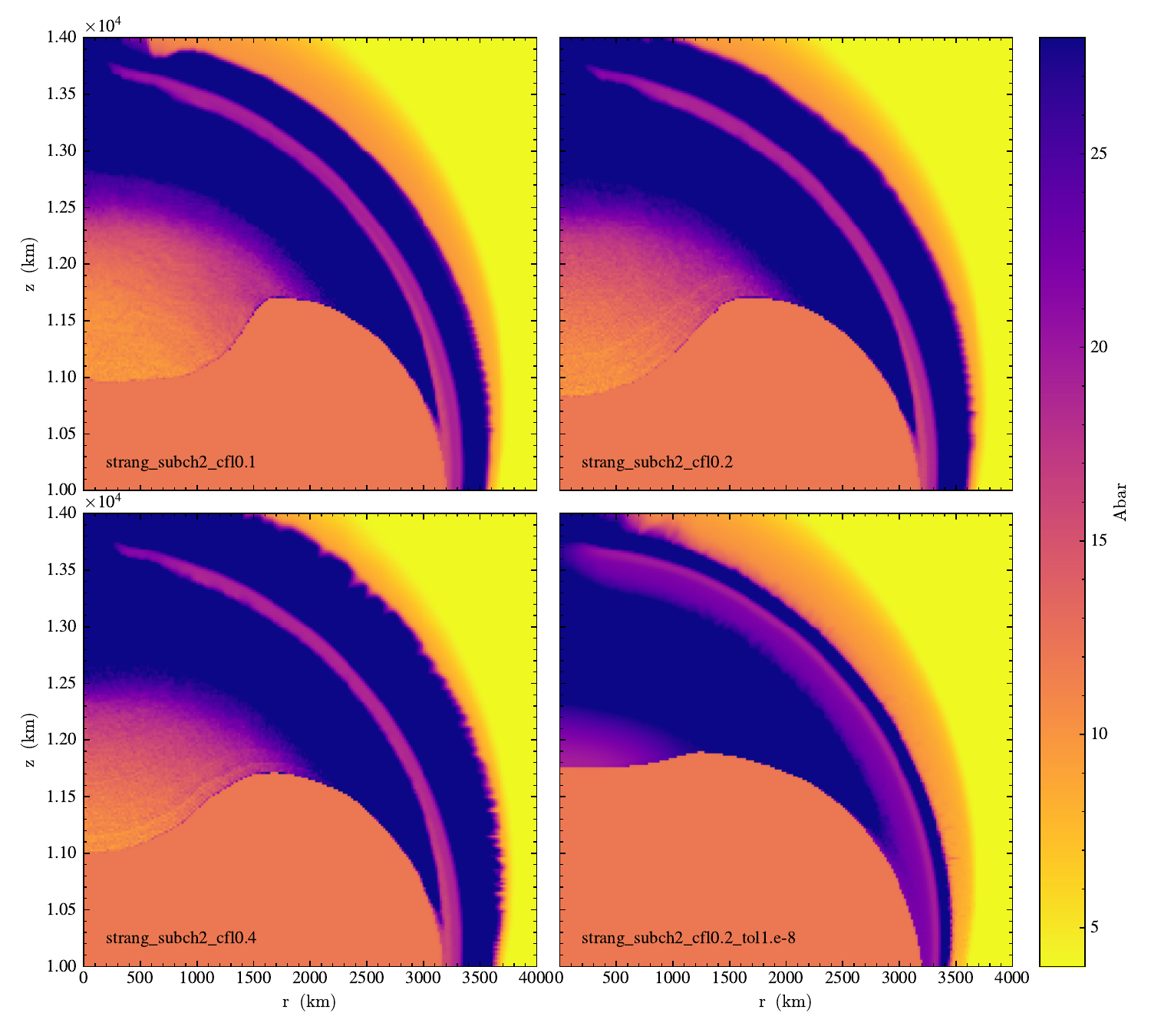}
\caption{\label{fig:strang_abar_summary} Comparison of Strang runs showing
the mean molecular weight behind the inward propagating
shock.}
\end{figure}

We next look at how the structure seen in
Figures~\ref{fig:subch_strang_cfl02} and \ref{fig:subch_sdc_cfl02}
varies with CFL number and network tolerance.  We focus just on the
region behind the inward propagating shock.
Figures~\ref{fig:strang_enuc_summary} and
\ref{fig:strang_abar_summary} show the nuclear energy generation rate
and mean molecular weight for all of the Strang runs.  We see that
changing the CFL number does little to affect the state behind the
shock---it continues to take on the mottled appearance characteristic
of not reaching a proper equilibrium.  For the case with tight tolerances, we
see that the state is greatly improved with a smoother gradient.  Furthermore, the inwardly
propagating carbon detonation reaches the same position as in the SDC runs.

\begin{figure}
\plotone{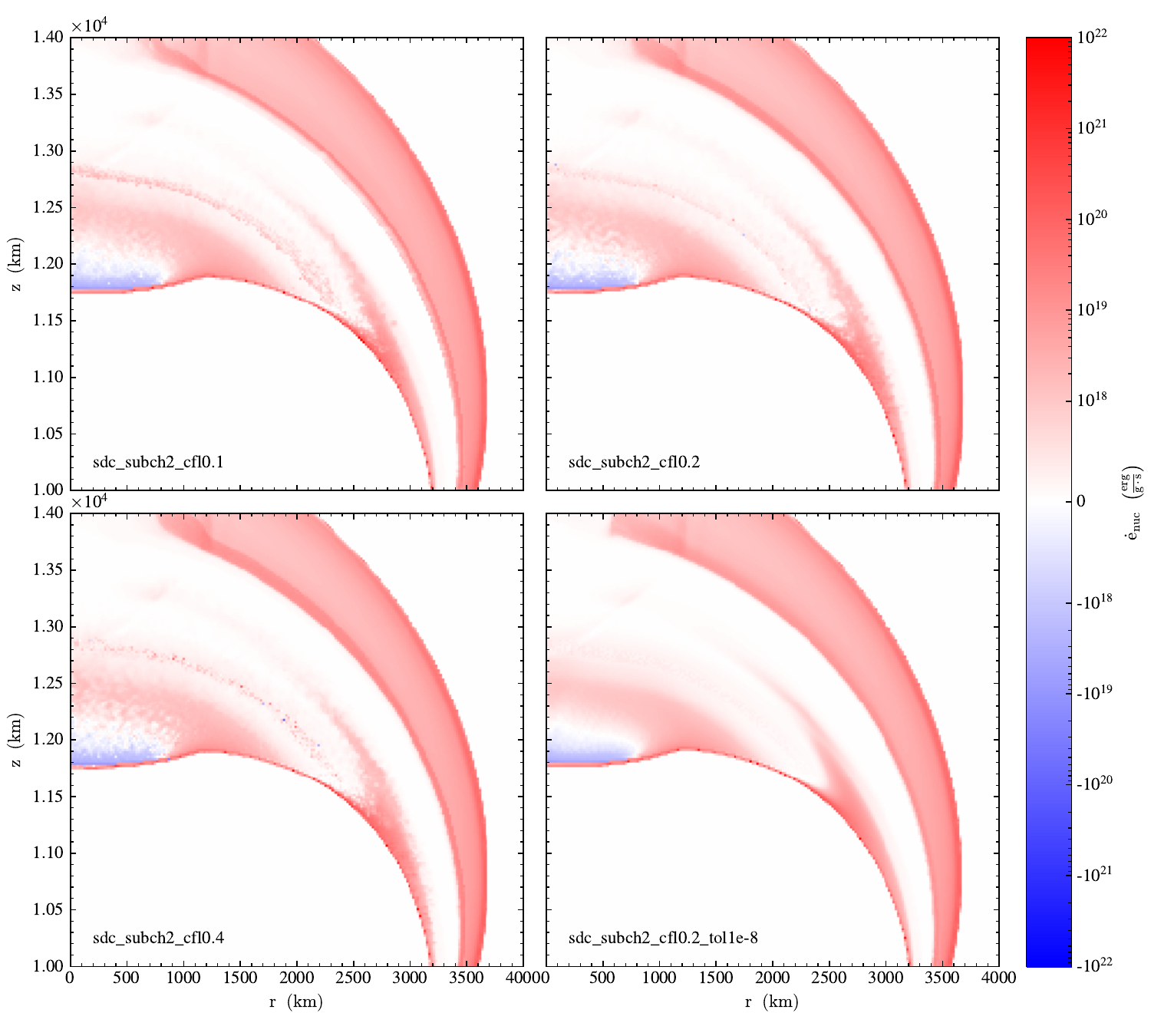}
\caption{\label{fig:sdc_enuc_summary} Comparison of simplified-SDC runs showing
the nuclear energy generation rate behind the inward propagating
shock.}
\end{figure}

\begin{figure}
\plotone{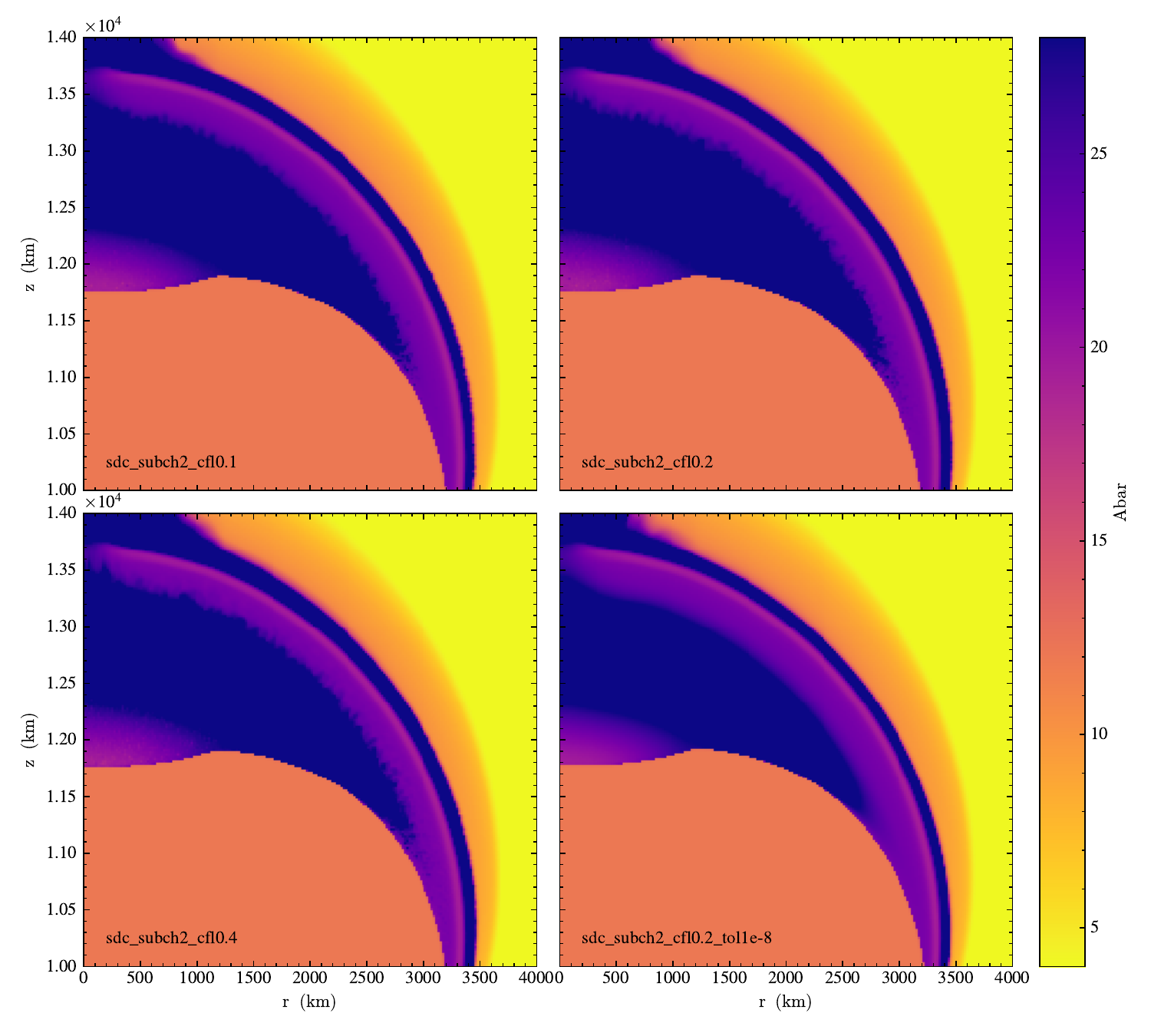}
\caption{\label{fig:sdc_abar_summary} Comparison of simplified-SDC runs showing
the mean molecular weight behind the inward propagating
shock.}
\end{figure}

The SDC comparisons are shown in Figures~\ref{fig:sdc_enuc_summary}
and \ref{fig:sdc_abar_summary}.  We see that the state behind the
shock looks smoother and more consistent regardless of the CFL number or integrator tolerances.
This demonstrates that the simplified-SDC integration algorithm is more
robust than Strang integration for this problem.

To quantify this effect, we next look at the total mass of some key
species.  This is often a desired diagnostic for supernova
calculations, since \isot{Ni}{56} powers the lightcurve.  Table
\ref{table:masses} shows the total mass of \isot{He}{4},
\isot{Si}{28}, and \isot{Ni}{56} for each of the simulations.  For the
Strang runs, there are large variations between the different
simulations while the SDC simulations all agree to within a few
percent.  Further, the Strang run with the tighter tolerances is in
good agreement with the masses found by the SDC runs.  This supports
the ideas described above that all of the SDC runs agree while only
the Strang run with the tighter tolerances is reliable.

\begin{deluxetable}{llll}
\tablecaption{\label{table:masses} Species masses.}
\tablehead{\colhead{simulation} & \colhead{$M_{\isotm{He}{4}}$} & \colhead{$M_{\isotm{Si}{28}}$} & \colhead{$M_{\isotm{Ni}{56}}$}}
\startdata
\multicolumn{4}{c}{Strang runs} \\
\hline
{\tt strang\_subch2\_cfl0.1}           & $1.059\times 10^{32}$  &  $3.804\times 10^{31}$  &  $1.092\times 10^{32}$ \\
{\tt strang\_subch2\_cfl0.2}           & $9.114\times 10^{31}$  &  $4.174\times 10^{31}$  &  $1.078\times 10^{32}$ \\
{\tt strang\_subch2\_cfl0.2\_tol1.e-8} & $7.222\times 10^{31}$  &  $8.623\times 10^{31}$  &  $4.041\times 10^{31}$ \\
{\tt strang\_subch2\_cfl0.4}           & $7.819\times 10^{31}$  &  $4.208\times 10^{31}$  &  $1.045\times 10^{32}$ \\
\hline
\multicolumn{4}{c}{SDC runs} \\
\hline
{\tt sdc\_subch2\_cfl0.1}           & $7.319\times 10^{31}$  &  $9.147\times 10^{31}$  &  $3.765\times 10^{31}$ \\
{\tt sdc\_subch2\_cfl0.2}           & $7.208\times 10^{31}$  &  $9.151\times 10^{31}$  &  $3.783\times 10^{31}$ \\
{\tt sdc\_subch2\_cfl0.2\_tol1.e-8} & $7.376\times 10^{31}$  &  $9.220\times 10^{31}$  &  $3.513\times 10^{31}$ \\
{\tt sdc\_subch2\_cfl0.4}           & $7.203\times 10^{31}$  &  $8.843\times 10^{31}$  &  $3.754\times 10^{31}$ \\
\hline
\enddata
\end{deluxetable}

\begin{figure}[t]
\plotone{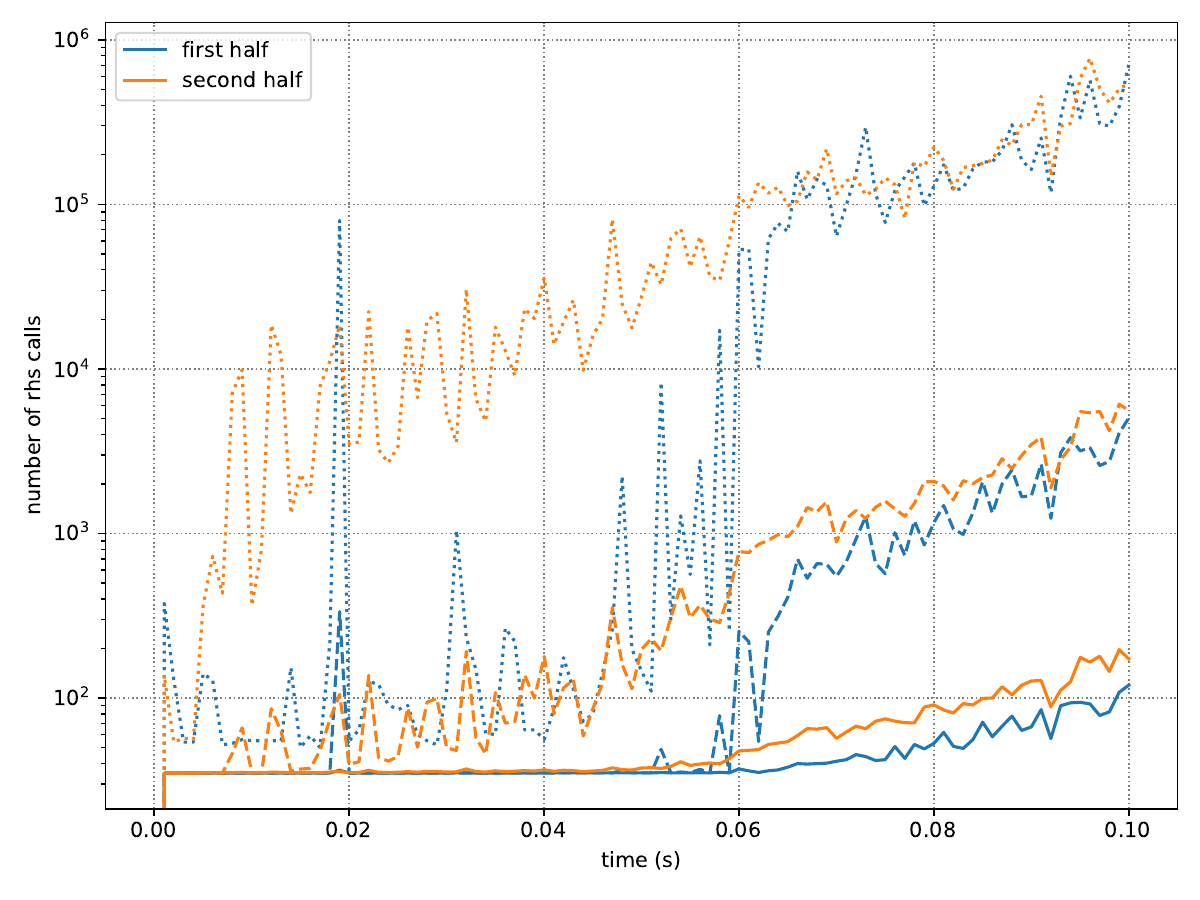}
\caption{\label{fig:strang_nrhs} Number of righthand side evaluations for {\tt strang\_subch2\_cfl0.2} showing the L1 norm/average (solid), L2 norm (dashed), and L-inf/maximum (dotted)
for both halves of the Strang integration.}
\end{figure}

\begin{figure}[t]
\plotone{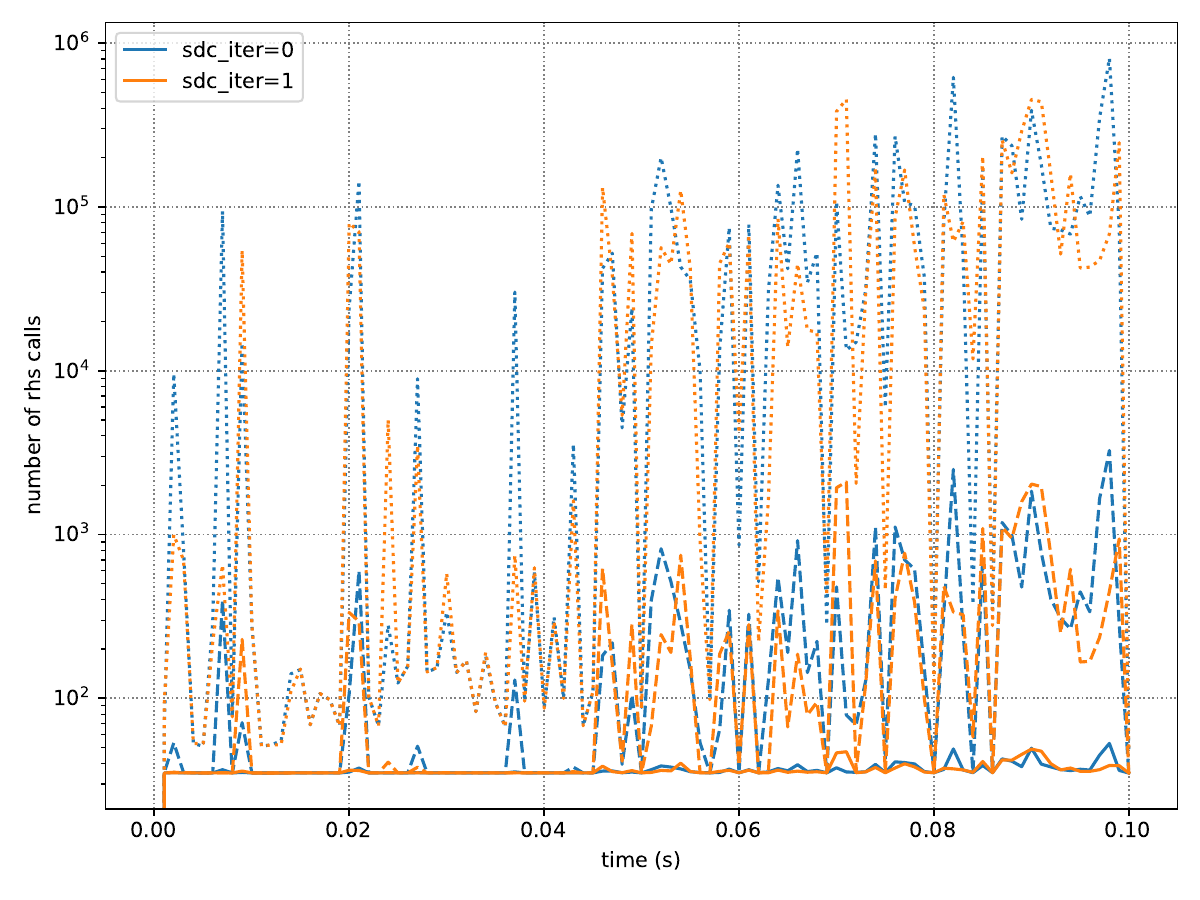}
\caption{\label{fig:sdc_nrhs} Number of righthand side evaluations for {\tt sdc\_subch2\_cfl0.2} showing the L1 norm/average (solid), L2 norm (dashed), and L-inf/maximum (dotted)
for 2 iterations of the simplified-SDC integration.}
\end{figure}

Finally, we can ask how many right-hand side evaluations each integration method required, as a measure of computational cost; when an integration uses many RHS evaluations, it often implies difficulty around equilibrium.  We stored this in the plotfiles every $0.001~\mathrm{s}$, and on the finest level, for zones that were burning, we compute the L1, L2, and L-inf norms of the number of right-hand side calls (the L1 norm is the same as the average and the L-inf is the same as the maximum here).  This includes the calls that were made to evaluate the numerical Jacobian.  Figure~\ref{fig:strang_nrhs} shows the results for the {\tt strang\_subch2\_cfl0.2} run.  We see a steady
increase in the cost of the reactions as time progresses.  Note that the L-inf norm curve is much higher than the other curves, but this is dominated by only a few zones right at the detonation.  The other norms give a better sense of the total work in reactions.
Figure~\ref{fig:sdc_nrhs} shows the corresponding plot for the simplified-SDC case ({\tt sdc\_subch2\_cfl0.2}).  While the L-inf
norm is of the same magnitude, it fluctuates much more.  However,
both the L1 and L2 norms are smaller, indicating that the integration does not need to work as hard with the simplified-SDC integration as compared to Strang.
%\MarginPar{AJN: add a note about the 1e-8 tol strang run (that gives same looking results as %SDC) probably is even MORE expensive?}

\section{Summary}

We presented a simplified spectral deferred corrections scheme for coupling
hydrodynamics and reactions.  We demonstrated that for a moderate-sized reaction
network, the simplified-SDC method compares well against Strang for the double
detonation problem---getting a better solution at looser reaction tolerances,
consistent results across a range of CFL numbers, and is more computationally efficient.
The benefits of this new
integration scheme are likely problem and network dependent, and we will explore
it on other science problems in future papers.

We have not focused on performance or load-balancing the reactions in
the tests that we ran, but for the simulations shown here,
the simplified-SDC method runs as fast (or faster) than the Strang splitting method
for the same tolerances and CFL number.
%, despite having to do the advection and reactions
%twice.
%\MarginPar{AJN: I would delete "despite having to to the advection and reactions twice".  %Strang already does reactions twice, but advection once.  But advection is cheap, right?}
We would expect similar behavior for any simulations that are
dominated by the cost of reactions.  If the results above are robust across
different problems and networks, then this suggests that we can run
the SDC simulation with larger CFL number and/or less strict reaction tolerances,
which would translate into a large performance boost.

This new time-integration method is freely available in \castro, and
we will continue to explore it on other problems.  Although these simulations
were run on CPUs, the entire simplified-SDC algorithm also runs on GPUs.
We are interested in massive star evolution with this method, where the
vigorous shell burning leading up to core-collapse would be an ideal
application for the simplified-SDC method.  In a follow-on paper, we
will show how to couple this new integration method to a table for
evaluating an NSE distribution in the iron core of a massive star.

Future work is to extend this methodology to MHD.  This is very
straightforward since the simplified-SDC method operates solely on the
thermodynamic state.  We will also explore how to couple radiation
hydrodynamics, where we need to do an implicit update for the
radiation energy.

\begin{acknowledgments} 
\castro\ is freely available at
\url{https://github.com/AMReX-Astro/Castro}.  All of the code and problem setups
used here are available in the git repo.  The double detonation problem is in
{\tt Castro/Exec/science/subchandra}.  The reaction network used
here is available at \url{https://github.com/AMReX-Astro/Microphysics} as the {\tt subch2} network.  The work at Stony Brook was supported by
DOE/Office of Nuclear Physics grant DE-FG02-87ER40317.  We thank Alice Harpole
for her contributions to the AMReX Astrophysics suite.  This material is based
upon work supported by the U.S. Department of Energy, Office of Science, Office
of Advanced Scientific Computing Research and Office of Nuclear Physics,
Scientific Discovery through Advanced Computing (SciDAC) program under Award
Number DE-SC0017955.  This research was supported by the Exascale Computing
Project (17-SC-20-SC), a collaborative effort of the U.S. Department of Energy
Office of Science and the National Nuclear Security Administration.  MR was supported
via an NSF REU grant to Stony Brook, NSF 1852143.  This
research used resources of the National Energy Research Scientific Computing
Center (NERSC), a U.S. Department of Energy Office of Science User Facility
located at Lawrence Berkeley National Laboratory, operated under Contract No.\
DE-AC02-05CH11231 using NERSC award NP-ERCAP0020354.
\end{acknowledgments}

\software{\amrex\ \citep{amrex_joss},
          \castro\ \citep{castro},
          GNU Compiler Collection (\url{https://gcc.gnu.org/}),
          Linux (\url{https://www.kernel.org}),
          matplotlib (\citealt{Hunter:2007},  \url{http://matplotlib.org/}),
          NetworkX \citep{networkx},
          NumPy \citep{numpy,numpy2},
          python (\url{https://www.python.org/}),
          \pynucastro\ \citep{pynucastro},
          pytest \citep{pytest},
          SymPy \citep{sympy},
          yt \citep{yt}
         }

\appendix

\section{Jacobian}

\label{sec:app:jacobian}

To solve the reaction system implicitly, the ODE solver needs the
Jacobian, $\partial \Rb/\partial \Uc^\prime$, where $\Uc^\prime =
(\rho X_k, \rho e)^\intercal$ is the subset of the conserved variables we are
integrating.  We follow the method of \cite{castro_sdc} and factor
this into two pieces,
\begin{equation}
\label{eq:factored_jac}
{\bf J} = \frac{\partial \Rb}{\partial {\bf w }} \frac{\partial {\bf w}}{\partial \Uc^\prime}.
\end{equation}
where the state ${\bf w}$ is chosen to match the set of variables used to evaluate the reaction rates.
Writing this out for two species, $X_\alpha$ and $X_\beta$, we have
\begin{equation}
\Uc^\prime = \left ( \begin{array}{c} \rho X_\alpha \\ \rho X_\beta \\ \rho e \end{array} \right )
\end{equation}
For interfacing with the reaction network, we use
\begin{equation}
{\bf w} = \left ( \begin{array}{c} X_\alpha \\  X_\beta \\ T \end{array} \right )
\end{equation}
Note: even though we are using $T$ here instead of $e$, we still do the overall ODE integration
in terms of $(\rho e)$, consistent with the Strang method described in \cite{strang_rnaas}.
%%  in that
%% we use $e$ instead of $T$ as the independent variable.  If the network gives derivatives in terms
%% of temperature, we can convert as:
%% \begin{equation}
%% \left . \frac{\partial \phi}{\partial T} \right |_\rho =
%%   \left . \frac{\partial \phi}{\partial e} \right |_\rho
%%   \left . \frac{\partial e}{\partial T} \right |_\rho =
%%   c_v \left . \frac{\partial \phi}{\partial e} \right |_\rho
%% \end{equation}
%% for some quantity $\phi(\rho, T)$.  Here we identified the specific heat
%% at constant volume, $c_v = \partial e/\partial T|_\rho$.

The Jacobian transformation $\partial \Uc^\prime/\partial {\bf w}$ is:
\begin{equation}
\frac{\partial \Uc^\prime}{\partial {\bf w}} = \left (
   \begin{array}{ccc}
       \rho & 0 & 0 \\
       0 & \rho & 0  \\
       \rho  e_{X_\alpha} & \rho e_{X_\beta} & \rho c_v \\
     \end{array}\right)
\end{equation}
where we use the following notation for compactness:
\begin{equation}
e_{X_k} = \dedXd
\end{equation}
and the specific heat at constant volume is
\begin{equation}
c_v = \left . \frac{\partial e}{\partial T} \right |_{\rho, X_k}
\end{equation}
We get the inverse
(computed via SymPy) as:
\begin{equation}
\renewcommand{\arraystretch}{1.5}
\frac{\partial {\bf w}}{\partial \Uc^\prime} = \left (
  \begin{array}{ccc}
   \frac{1}{\rho} & 0 & 0 \\
   0 & \frac{1}{\rho} & 0 \\
    -\frac{e_{X_\alpha}}{\rho c_v} & -\frac{e_{X_\beta}}{\rho c_v} & \frac{1}{\rho c_v} \\
   \end{array}\right)
\renewcommand{\arraystretch}{1}
\end{equation}

The reaction vector is
\begin{equation}
\Rb(\Uc^\prime) = \left (  \begin{array}{c} \rho \omegadot_\alpha \\ \rho \omegadot_\beta \\ \rho \Sdot \end{array} \right )
\end{equation}
We take all the quantities to be functions of $\rho$, $e$, and $X_k$,
but since $\rho$ doesn't change from the reactions (the reactive
source of the continuity equation is zero), it is held constant in
these derivatives.  The Jacobian is computed as $\partial \Rb/\partial
{\bf w}$:
\begin{equation}
\renewcommand{\arraystretch}{1.5}
\frac{\partial \Rb}{\partial {\bf w}} = \left (
  \begin{array}{ccc}
     \rho \frac{\partial \omegadot_\alpha}{\partial X_\alpha} &
     \rho \frac{\partial \omegadot_\alpha}{\partial X_\beta} &
     \rho \frac{\partial \omegadot_\alpha}{\partial T} \\
     \rho \frac{\partial \omegadot_\beta}{\partial X_\alpha} &
     \rho \frac{\partial \omegadot_\beta}{\partial X_\beta} &
     \rho \frac{\partial \omegadot_\beta}{\partial T} \\
     \rho \frac{\partial \Sdot}{\partial X_\alpha} &
     \rho \frac{\partial \Sdot}{\partial X_\beta} &
     \rho \frac{\partial \Sdot}{\partial T} \\
  \end{array}
  \right )
\renewcommand{\arraystretch}{1}
\end{equation}

The final Jacobian is found by multiplying these two:
\begin{equation}
\renewcommand{\arraystretch}{1.5}
\frac{\partial \Rb}{\partial \Uc^\prime} = \left (
  \begin{array}{ccc}
    \frac{\partial \omegadot_\alpha}{\partial X_\alpha} - \frac{e_{X_\alpha}}{c_v} \frac{\partial \omegadot_\alpha}{\partial T} &
    \frac{\partial \omegadot_\alpha}{\partial X_\beta} - \frac{e_{X_\beta}}{c_v} \frac{\partial \omegadot_\alpha}{\partial T} &
    \frac{1}{c_v} \frac{\partial \omegadot_\alpha}{\partial T} \\
    \frac{\partial \omegadot_\beta}{\partial X_\alpha} - \frac{e_{X_\alpha}}{c_v} \frac{\partial \omegadot_\beta}{\partial T} &
    \frac{\partial \omegadot_\beta}{\partial X_\beta} - \frac{e_{X_\beta}}{c_v} \frac{\partial \omegadot_\beta}{\partial T} &
    \frac{1}{c_v} \frac{\partial \omegadot_\beta}{\partial T} \\
     \frac{\partial \Sdot}{\partial X_\alpha} -  \frac{e_{X_\alpha}}{c_v} \frac{\partial \Sdot}{\partial T} &
     \frac{\partial \Sdot}{\partial X_\beta} -  \frac{e_{X_\beta}}{c_v} \frac{\partial \Sdot}{\partial T} &
     \frac{1}{c_v} \frac{\partial \Sdot}{\partial T} \\
  \end{array}
  \right )
\renewcommand{\arraystretch}{1}
\end{equation}
We note that the form of these entries is the same as one would arrive at if you start with the rates expressed as
$\omega_k(\rho, T(\rho, X_j, e), X_j)$ and recognize that constant $e$ implies that
\begin{equation}
\left . \frac{\partial T}{\partial X_k} \right |_{\rho, e,X_{j,j\ne k}} = - \frac{e_{X_k}}{c_v}
\end{equation}

Finally we note that one implication of this formulation is that the Jacobian is no longer sparse.  Future
work will explore alternate formulations and approximate Jacobians.

% While VODE can compute the entire Jacobian, ${\bf J}$ numerically via
% differencing, we found that does not give reliable results.  Instead,
% we compute compute the derivatives with respect to $X_k$ and $T$
% one-sided differencing, following the algorithm in \cite{lsode} to
% minimize numerical noise.  We then use the equation of state to
% compute $e_{X_k}$ and $c_v$ and construct the entries of the final
% Jacobian.

\section{Reaction Network}
\label{sec:app:reactionnet}
We build a reaction network that approximates an $\alpha$-chain,
including all $\isotm{He}{4}$, and all of the $\alpha$ nuclei from
$\isotm{C}{12}$ to $\isotm{Ni}{56}$.  We want to capture both the
$(\alpha,\gamma)$ and $(\alpha,p)(p,\gamma)$ links, so we include the
intermediate nuclei from the $(\alpha,p)$ rate instead of
approximating these links assuming proton equilibriation.  Following
\citet{shenbildsten}, we include the sequence
$\isotm{N}{14}(\alpha,\gamma)\isotm{F}{18}(\alpha, p)\isotm{Ne}{21}$,
which produce the protons that can allow for $\isotm{C}{12}(p,
\gamma)\isotm{N}{13}(\alpha, p)\isotm{O}{16}$ to compete with
$\isotm{C}{12}(\alpha,\gamma)\isotm{O}{16}$.  We add $\isotm{Na}{22}$
to link the $\isotm{Ne}{21}$ produced back to the $\alpha$-chain.  We
grab all of the ReacLib \citep{reaclib} reaction rates linking these
nuclei using \pynucastro~\citep{pynucastro}.  We are missing one rate
each for $\isotm{C}{12} + \isotm{C}{12}$, $\isotm{C}{12} +
\isotm{O}{16}$, and $\isotm{O}{16} + \isotm{O}{16}$.  These are
the sequences $\isotm{C}{12}(\isotm{C}{12}, \mathrm{n})\isotm{Mg}{23}(n, \gamma)\isotm{Mg}{24}$,
$\isotm{O}{16}(\isotm{C}{12}, \mathrm{n})\isotm{Si}{27}(n, \gamma)\isotm{Si}{28}$, and
$\isotm{O}{16}(\isotm{O}{16}, \mathrm{n})\isotm{S}{31}(n, \gamma)\isotm{S}{32}$, involving neutron
production followed by a capture back to one of our $\alpha$-chain
nuclei.  These are the only rates we approximate here, assuming
that the subsequent neutron capture is instantaneous.

\begin{figure}[t]
\centering
\plotone{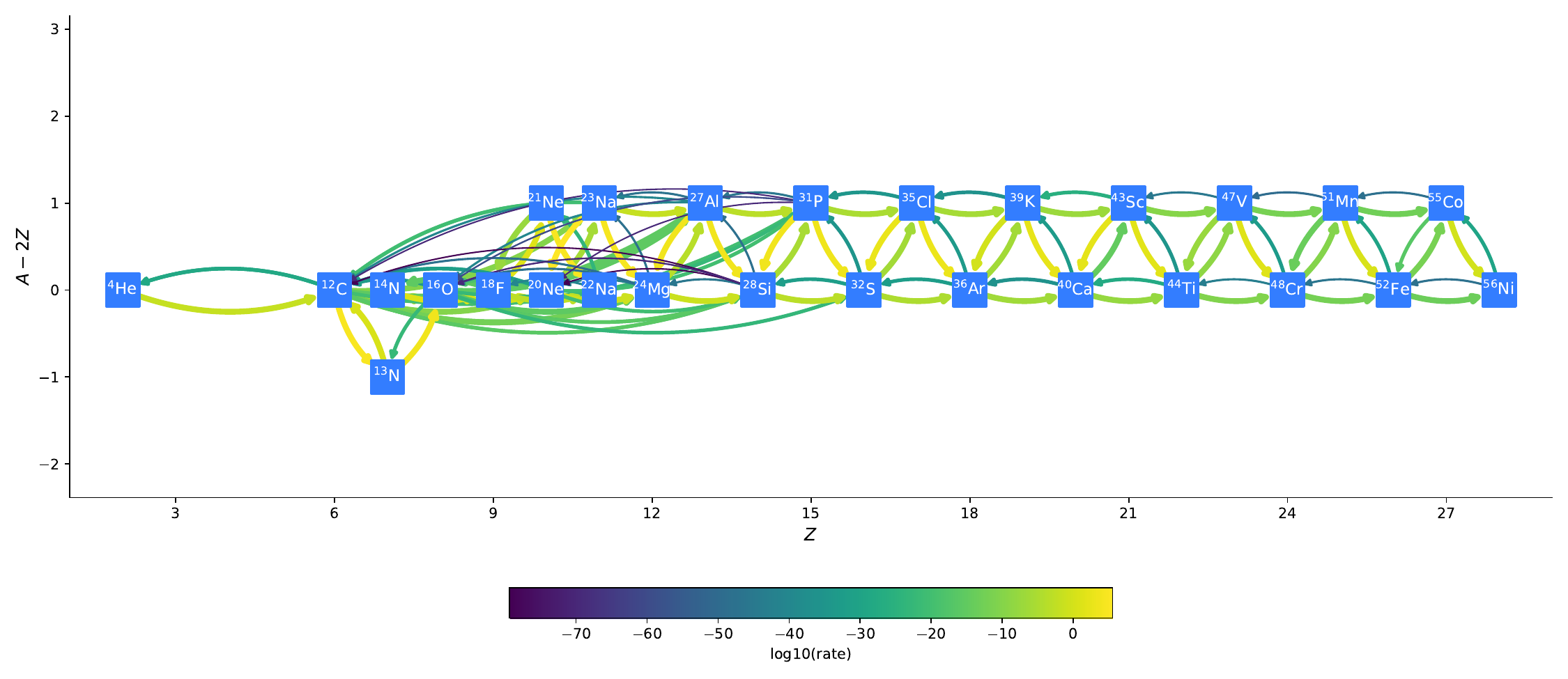}
\caption{\label{fig:subch_network} Reaction network for the double detonation problem.}
\end{figure}

Our final set of 28 nuclei are ${}^{1}\mathrm{H}$, ${}^{4}\mathrm{He}$,
${}^{12}\mathrm{C}$, ${}^{13}\mathrm{N}$, ${}^{14}\mathrm{N}$,
${}^{16}\mathrm{O}$, ${}^{18}\mathrm{F}$, ${}^{20}\mathrm{Ne}$,
${}^{21}\mathrm{Ne}$, ${}^{22}\mathrm{Na}$, ${}^{23}\mathrm{Na}$,
${}^{24}\mathrm{Mg}$, ${}^{27}\mathrm{Al}$, ${}^{28}\mathrm{Si}$,
${}^{31}\mathrm{P}$, ${}^{32}\mathrm{S}$, ${}^{35}\mathrm{Cl}$,
${}^{36}\mathrm{Ar}$, ${}^{39}\mathrm{K}$, ${}^{40}\mathrm{Ca}$,
${}^{43}\mathrm{Sc}$, ${}^{44}\mathrm{Ti}$, ${}^{47}\mathrm{V}$,
${}^{48}\mathrm{Cr}$, ${}^{51}\mathrm{Mn}$, ${}^{52}\mathrm{Fe}$,
${}^{55}\mathrm{Co}$, ${}^{56}\mathrm{Ni}$.  The 107 rates linking them are
given in table~\ref{table:rates}, with the forward rates having a positive $Q$
value and the reverse having a negative $Q$ value.  We do not recompute the reverse
rates via detailed balance or correct them with high-temperature partition functions---this will be considered in a future science-focused paper.  We add screening to the
rates using the prescription from \citet{graboske:1973,alastuey:1978,itoh:1979}.
The C++ code for the network is output by \pynucastro\ directly for \castro.  A
graphical overview of the network showing the links between the nuclei is shown
in figure~\ref{fig:subch_network}.

\startlongtable
\begin{deluxetable}{ll}
\tablecaption{\label{table:rates} Reaction rates included in our network.}
\tablehead{\colhead{forward reaction} & \colhead{reverse reaction}}
\startdata
${}^{4}\mathrm{He} + {}^{4}\mathrm{He} + {}^{4}\mathrm{He} \rightarrow {}^{12}\mathrm{C}$ &
  ${}^{12}\mathrm{C} \rightarrow {}^{4}\mathrm{He} + {}^{4}\mathrm{He} + {}^{4}\mathrm{He}$ \\
${}^{12}\mathrm{C} + {}^{1}\mathrm{H} \rightarrow {}^{13}\mathrm{N}$ &
  ${}^{13}\mathrm{N} \rightarrow {}^{1}\mathrm{H} + {}^{12}\mathrm{C}$ \\
${}^{12}\mathrm{C} + {}^{4}\mathrm{He} \rightarrow {}^{16}\mathrm{O}$ &
  ${}^{16}\mathrm{O} \rightarrow {}^{4}\mathrm{He} + {}^{12}\mathrm{C}$ \\
${}^{12}\mathrm{C} + {}^{12}\mathrm{C} \rightarrow {}^{4}\mathrm{He} + {}^{20}\mathrm{Ne}$ &
  ${}^{20}\mathrm{Ne} + {}^{4}\mathrm{He} \rightarrow {}^{12}\mathrm{C} + {}^{12}\mathrm{C}$ \\
${}^{12}\mathrm{C} + {}^{12}\mathrm{C} \rightarrow {}^{1}\mathrm{H} + {}^{23}\mathrm{Na}$ &
  ${}^{23}\mathrm{Na} + {}^{1}\mathrm{H} \rightarrow {}^{12}\mathrm{C} + {}^{12}\mathrm{C}$ \\
${}^{12}\mathrm{C} + {}^{12}\mathrm{C} \rightarrow {}^{24}\mathrm{Mg}$ &
                                         \\
${}^{13}\mathrm{N} + {}^{4}\mathrm{He} \rightarrow {}^{1}\mathrm{H} + {}^{16}\mathrm{O}$ &
  ${}^{16}\mathrm{O} + {}^{1}\mathrm{H} \rightarrow {}^{4}\mathrm{He} + {}^{13}\mathrm{N}$ \\
${}^{14}\mathrm{N} + {}^{4}\mathrm{He} \rightarrow {}^{18}\mathrm{F}$ &
  ${}^{18}\mathrm{F} \rightarrow {}^{4}\mathrm{He} + {}^{14}\mathrm{N}$ \\
${}^{16}\mathrm{O} + {}^{4}\mathrm{He} \rightarrow {}^{20}\mathrm{Ne}$ &
  ${}^{20}\mathrm{Ne} \rightarrow {}^{4}\mathrm{He} + {}^{16}\mathrm{O}$ \\
${}^{16}\mathrm{O} + {}^{12}\mathrm{C} \rightarrow {}^{4}\mathrm{He} + {}^{24}\mathrm{Mg}$ &
  ${}^{24}\mathrm{Mg} + {}^{4}\mathrm{He} \rightarrow {}^{12}\mathrm{C} + {}^{16}\mathrm{O}$ \\
${}^{16}\mathrm{O} + {}^{12}\mathrm{C} \rightarrow {}^{1}\mathrm{H} + {}^{27}\mathrm{Al}$ &
  ${}^{27}\mathrm{Al} + {}^{1}\mathrm{H} \rightarrow {}^{12}\mathrm{C} + {}^{16}\mathrm{O}$ \\
${}^{16}\mathrm{O} + {}^{12}\mathrm{C} \rightarrow {}^{28}\mathrm{Si}$ &
                                         \\
${}^{16}\mathrm{O} + {}^{16}\mathrm{O} \rightarrow {}^{4}\mathrm{He} + {}^{28}\mathrm{Si}$ &
  ${}^{28}\mathrm{Si} + {}^{4}\mathrm{He} \rightarrow {}^{16}\mathrm{O} + {}^{16}\mathrm{O}$ \\
${}^{16}\mathrm{O} + {}^{16}\mathrm{O} \rightarrow {}^{1}\mathrm{H} + {}^{31}\mathrm{P}$ &
  ${}^{31}\mathrm{P} + {}^{1}\mathrm{H} \rightarrow {}^{16}\mathrm{O} + {}^{16}\mathrm{O}$ \\
${}^{16}\mathrm{O} + {}^{16}\mathrm{O} \rightarrow {}^{32}\mathrm{S}$ &
                                         \\
${}^{18}\mathrm{F} + {}^{4}\mathrm{He} \rightarrow {}^{1}\mathrm{H} + {}^{21}\mathrm{Ne}$ &
  ${}^{21}\mathrm{Ne} + {}^{1}\mathrm{H} \rightarrow {}^{4}\mathrm{He} + {}^{18}\mathrm{F}$ \\
${}^{18}\mathrm{F} + {}^{4}\mathrm{He} \rightarrow {}^{22}\mathrm{Na}$ &
  ${}^{22}\mathrm{Na} \rightarrow {}^{4}\mathrm{He} + {}^{18}\mathrm{F}$ \\
${}^{20}\mathrm{Ne} + {}^{4}\mathrm{He} \rightarrow {}^{24}\mathrm{Mg}$ &
  ${}^{24}\mathrm{Mg} \rightarrow {}^{4}\mathrm{He} + {}^{20}\mathrm{Ne}$ \\
${}^{20}\mathrm{Ne} + {}^{12}\mathrm{C} \rightarrow {}^{4}\mathrm{He} + {}^{28}\mathrm{Si}$ &
  ${}^{28}\mathrm{Si} + {}^{4}\mathrm{He} \rightarrow {}^{12}\mathrm{C} + {}^{20}\mathrm{Ne}$ \\
${}^{20}\mathrm{Ne} + {}^{12}\mathrm{C} \rightarrow {}^{1}\mathrm{H} + {}^{31}\mathrm{P}$ &
  ${}^{31}\mathrm{P} + {}^{1}\mathrm{H} \rightarrow {}^{12}\mathrm{C} + {}^{20}\mathrm{Ne}$ \\
${}^{21}\mathrm{Ne} + {}^{1}\mathrm{H} \rightarrow {}^{22}\mathrm{Na}$ &
  ${}^{22}\mathrm{Na} \rightarrow {}^{1}\mathrm{H} + {}^{21}\mathrm{Ne}$ \\
${}^{23}\mathrm{Na} + {}^{1}\mathrm{H} \rightarrow {}^{4}\mathrm{He} + {}^{20}\mathrm{Ne}$ &
  ${}^{20}\mathrm{Ne} + {}^{4}\mathrm{He} \rightarrow {}^{1}\mathrm{H} + {}^{23}\mathrm{Na}$ \\
${}^{23}\mathrm{Na} + {}^{1}\mathrm{H} \rightarrow {}^{24}\mathrm{Mg}$ &
  ${}^{24}\mathrm{Mg} \rightarrow {}^{1}\mathrm{H} + {}^{23}\mathrm{Na}$ \\
${}^{23}\mathrm{Na} + {}^{4}\mathrm{He} \rightarrow {}^{27}\mathrm{Al}$ &
  ${}^{27}\mathrm{Al} \rightarrow {}^{4}\mathrm{He} + {}^{23}\mathrm{Na}$ \\
${}^{24}\mathrm{Mg} + {}^{4}\mathrm{He} \rightarrow {}^{28}\mathrm{Si}$ &
  ${}^{28}\mathrm{Si} \rightarrow {}^{4}\mathrm{He} + {}^{24}\mathrm{Mg}$ \\
${}^{27}\mathrm{Al} + {}^{1}\mathrm{H} \rightarrow {}^{4}\mathrm{He} + {}^{24}\mathrm{Mg}$ &
  ${}^{24}\mathrm{Mg} + {}^{4}\mathrm{He} \rightarrow {}^{1}\mathrm{H} + {}^{27}\mathrm{Al}$ \\
${}^{27}\mathrm{Al} + {}^{1}\mathrm{H} \rightarrow {}^{28}\mathrm{Si}$ &
  ${}^{28}\mathrm{Si} \rightarrow {}^{1}\mathrm{H} + {}^{27}\mathrm{Al}$ \\
${}^{27}\mathrm{Al} + {}^{4}\mathrm{He} \rightarrow {}^{31}\mathrm{P}$ &
  ${}^{31}\mathrm{P} \rightarrow {}^{4}\mathrm{He} + {}^{27}\mathrm{Al}$ \\
${}^{28}\mathrm{Si} + {}^{4}\mathrm{He} \rightarrow {}^{32}\mathrm{S}$ &
  ${}^{32}\mathrm{S} \rightarrow {}^{4}\mathrm{He} + {}^{28}\mathrm{Si}$ \\
${}^{31}\mathrm{P} + {}^{1}\mathrm{H} \rightarrow {}^{4}\mathrm{He} + {}^{28}\mathrm{Si}$ &
  ${}^{28}\mathrm{Si} + {}^{4}\mathrm{He} \rightarrow {}^{1}\mathrm{H} + {}^{31}\mathrm{P}$ \\
${}^{31}\mathrm{P} + {}^{1}\mathrm{H} \rightarrow {}^{32}\mathrm{S}$ &
  ${}^{32}\mathrm{S} \rightarrow {}^{1}\mathrm{H} + {}^{31}\mathrm{P}$ \\
${}^{31}\mathrm{P} + {}^{4}\mathrm{He} \rightarrow {}^{35}\mathrm{Cl}$ &
  ${}^{35}\mathrm{Cl} \rightarrow {}^{4}\mathrm{He} + {}^{31}\mathrm{P}$ \\
${}^{32}\mathrm{S} + {}^{4}\mathrm{He} \rightarrow {}^{36}\mathrm{Ar}$ &
  ${}^{36}\mathrm{Ar} \rightarrow {}^{4}\mathrm{He} + {}^{32}\mathrm{S}$ \\
${}^{35}\mathrm{Cl} + {}^{1}\mathrm{H} \rightarrow {}^{4}\mathrm{He} + {}^{32}\mathrm{S}$ &
  ${}^{32}\mathrm{S} + {}^{4}\mathrm{He} \rightarrow {}^{1}\mathrm{H} + {}^{35}\mathrm{Cl}$ \\
${}^{35}\mathrm{Cl} + {}^{1}\mathrm{H} \rightarrow {}^{36}\mathrm{Ar}$ &
  ${}^{36}\mathrm{Ar} \rightarrow {}^{1}\mathrm{H} + {}^{35}\mathrm{Cl}$ \\
${}^{35}\mathrm{Cl} + {}^{4}\mathrm{He} \rightarrow {}^{39}\mathrm{K}$ &
  ${}^{39}\mathrm{K} \rightarrow {}^{4}\mathrm{He} + {}^{35}\mathrm{Cl}$ \\
${}^{36}\mathrm{Ar} + {}^{4}\mathrm{He} \rightarrow {}^{40}\mathrm{Ca}$ &
  ${}^{40}\mathrm{Ca} \rightarrow {}^{4}\mathrm{He} + {}^{36}\mathrm{Ar}$ \\
${}^{39}\mathrm{K} + {}^{1}\mathrm{H} \rightarrow {}^{4}\mathrm{He} + {}^{36}\mathrm{Ar}$ &
  ${}^{36}\mathrm{Ar} + {}^{4}\mathrm{He} \rightarrow {}^{1}\mathrm{H} + {}^{39}\mathrm{K}$ \\
${}^{39}\mathrm{K} + {}^{1}\mathrm{H} \rightarrow {}^{40}\mathrm{Ca}$ &
  ${}^{40}\mathrm{Ca} \rightarrow {}^{1}\mathrm{H} + {}^{39}\mathrm{K}$ \\
${}^{39}\mathrm{K} + {}^{4}\mathrm{He} \rightarrow {}^{43}\mathrm{Sc}$ &
  ${}^{43}\mathrm{Sc} \rightarrow {}^{4}\mathrm{He} + {}^{39}\mathrm{K}$ \\
${}^{40}\mathrm{Ca} + {}^{4}\mathrm{He} \rightarrow {}^{44}\mathrm{Ti}$ &
  ${}^{44}\mathrm{Ti} \rightarrow {}^{4}\mathrm{He} + {}^{40}\mathrm{Ca}$ \\
${}^{43}\mathrm{Sc} + {}^{1}\mathrm{H} \rightarrow {}^{4}\mathrm{He} + {}^{40}\mathrm{Ca}$ &
  ${}^{40}\mathrm{Ca} + {}^{4}\mathrm{He} \rightarrow {}^{1}\mathrm{H} + {}^{43}\mathrm{Sc}$ \\
${}^{43}\mathrm{Sc} + {}^{1}\mathrm{H} \rightarrow {}^{44}\mathrm{Ti}$ &
  ${}^{44}\mathrm{Ti} \rightarrow {}^{1}\mathrm{H} + {}^{43}\mathrm{Sc}$ \\
${}^{43}\mathrm{Sc} + {}^{4}\mathrm{He} \rightarrow {}^{47}\mathrm{V}$ &
  ${}^{47}\mathrm{V} \rightarrow {}^{4}\mathrm{He} + {}^{43}\mathrm{Sc}$ \\
${}^{44}\mathrm{Ti} + {}^{4}\mathrm{He} \rightarrow {}^{48}\mathrm{Cr}$ &
  ${}^{48}\mathrm{Cr} \rightarrow {}^{4}\mathrm{He} + {}^{44}\mathrm{Ti}$ \\
${}^{47}\mathrm{V} + {}^{1}\mathrm{H} \rightarrow {}^{4}\mathrm{He} + {}^{44}\mathrm{Ti}$ &
  ${}^{44}\mathrm{Ti} + {}^{4}\mathrm{He} \rightarrow {}^{1}\mathrm{H} + {}^{47}\mathrm{V}$ \\
${}^{47}\mathrm{V} + {}^{1}\mathrm{H} \rightarrow {}^{48}\mathrm{Cr}$ &
  ${}^{48}\mathrm{Cr} \rightarrow {}^{1}\mathrm{H} + {}^{47}\mathrm{V}$ \\
${}^{47}\mathrm{V} + {}^{4}\mathrm{He} \rightarrow {}^{51}\mathrm{Mn}$ &
  ${}^{51}\mathrm{Mn} \rightarrow {}^{4}\mathrm{He} + {}^{47}\mathrm{V}$ \\
${}^{48}\mathrm{Cr} + {}^{4}\mathrm{He} \rightarrow {}^{1}\mathrm{H} + {}^{51}\mathrm{Mn}$ &
  ${}^{51}\mathrm{Mn} + {}^{1}\mathrm{H} \rightarrow {}^{4}\mathrm{He} + {}^{48}\mathrm{Cr}$ \\
${}^{48}\mathrm{Cr} + {}^{4}\mathrm{He} \rightarrow {}^{52}\mathrm{Fe}$ &
  ${}^{52}\mathrm{Fe} \rightarrow {}^{4}\mathrm{He} + {}^{48}\mathrm{Cr}$ \\
${}^{51}\mathrm{Mn} + {}^{1}\mathrm{H} \rightarrow {}^{52}\mathrm{Fe}$ &
  ${}^{52}\mathrm{Fe} \rightarrow {}^{1}\mathrm{H} + {}^{51}\mathrm{Mn}$ \\
${}^{51}\mathrm{Mn} + {}^{4}\mathrm{He} \rightarrow {}^{55}\mathrm{Co}$ &
  ${}^{55}\mathrm{Co} \rightarrow {}^{4}\mathrm{He} + {}^{51}\mathrm{Mn}$ \\
${}^{52}\mathrm{Fe} + {}^{4}\mathrm{He} \rightarrow {}^{1}\mathrm{H} + {}^{55}\mathrm{Co}$ &
  ${}^{55}\mathrm{Co} + {}^{1}\mathrm{H} \rightarrow {}^{4}\mathrm{He} + {}^{52}\mathrm{Fe}$ \\
${}^{52}\mathrm{Fe} + {}^{4}\mathrm{He} \rightarrow {}^{56}\mathrm{Ni}$ &
  ${}^{56}\mathrm{Ni} \rightarrow {}^{4}\mathrm{He} + {}^{52}\mathrm{Fe}$ \\
${}^{55}\mathrm{Co} + {}^{1}\mathrm{H} \rightarrow {}^{56}\mathrm{Ni}$ &
  ${}^{56}\mathrm{Ni} \rightarrow {}^{1}\mathrm{H} + {}^{55}\mathrm{Co}$ \\
\enddata
\end{deluxetable}

%======================================================================
% References
%======================================================================

\bibliographystyle{aasjournal}
\bibliography{ws}

\end{document}